\documentclass[fleqn,usenatbib]{mnras}

\usepackage[T1]{fontenc}

\usepackage{xcolor}
\DeclareRobustCommand{\VAN}[3]{#2}
\let\VANthebibliography\thebibliography
\def\thebibliography{\DeclareRobustCommand{\VAN}[3]{##3}\VANthebibliography}
\definecolor{amethyst}{rgb}{0.6, 0.4, 0.8}
\definecolor{ao(english)}{rgb}{0.0, 0.5, 0.0}
\definecolor{cadmiumgreen}{rgb}{0.0, 0.42, 0.24}
\definecolor{grey}{rgb}{0.43, 0.5, 0.5}
\usepackage{graphicx}	% Including figure files
\usepackage{amsmath}	% Advanced maths commands
\usepackage{amssymb}	% Extra maths symbols
\usepackage{xspace}     % ---- xspace for \ion, etc. 
\usepackage{multirow}

\usepackage{newtxtext,newtxmath}
\DeclareSymbolFont{CMletters}{OML}{cmm}{m}{it}
\DeclareMathSymbol{\nu}{\mathord}{CMletters}{23}
\DeclareMathSymbol{v}{\mathord}{CMletters}{`v}

\DeclareRobustCommand{\ion}[2]{%
\relax\ifmmode
\ifx\testbx\f@series
{\mathbf{#1\,\mathsc{#2}}}\else
{\mathrm{#1\,\mathsc{#2}}}\fi
\else\textup{#1\,{\mdseries\textsc{#2}}}%
\fi}

\newcommand{\HI}{{\ion{H}{I}}\xspace}
\newcommand{\HII}{{\ion{H}{II}}\xspace}

\renewcommand{\d}[1]{\ensuremath{\operatorname{d}\!{#1}}}

\newcommand{\cf}{\mbox{cf.\,}}

\newcommand{\Fig}[1]{Fig.~#1}

\title[Transfer of 21-cm line through HII cavities]{
Radiative transfer of 21-cm line through ionised cavities \\ 
  in an expanding universe}

\author[Wu, Han and Chan] 
{Kinwah Wu$^{1}$\thanks{E-mail: kinwah.wu@ucl.ac.uk (KW), 
qin.han.21@ucl.ac.uk (QH), jyhchan@cita.utoronto.ca (JYHC)},  
Qin Han$^{1\star}$  and 
Jennifer Y. H. Chan$^{2,3,4}$  
\\
$^1$Mullard Space Science Laboratory, University College London, 
    Holmbury St Mary, Surrey, RH5 6NT, UK \\ 
$^2$Canadian Institute for Theoretical Astrophysics, University of Toronto, 60 St George St, Toronto, ON M5S 3H8, Canada     \\ 
$^3$Faculty of Arts and Science, University of Toronto, 100 St George St, Toronto, ON M5S 3G3, Canada \\
$^4$Dunlap Institute for Astronomy and Astrophysics, University of Toronto, 50 St George St, Toronto, ON M5S 3H4, Canada  
}

\date{Accepted 2024 May 17. Received 2024 May 17; in original form 2023 September 26}

\pubyear{2024}

\begin{document}
\label{firstpage}
\pagerange{\pageref{firstpage}--\pageref{lastpage}}
\maketitle

\begin{abstract} 
The optical depth parameterisation is typically used to study the 21-cm signals associated with the properties of the neutral hydrogen (\HI) gas and the ionisation morphology during the Epoch of Reionisation (EoR), without solving the radiative transfer equation. To assess the uncertainties resulting from this simplification, we conduct explicit radiative transfer calculations using the cosmological 21-cm radiative transfer (C21LRT) code and examine the imprints of ionisation structures on the 21-cm spectrum.    
We consider 
  a globally averaged reionisation history 
  and implement fully ionised cavities (\HII bubbles) 
  of diameters $d$ ranging from 0.01 Mpc to 10 Mpc  
  at epochs  
  within the emission and the absorption regimes 
  of the 21-cm global signal. 
The single-ray C21LRT calculations 
  show that the shape of the imprinted spectral features are primarily determined by $d$ and the 21-cm line profile, 
  which is parametrised by the turbulent velocity of the \HI gas. 
It reveals 
  the spectral features tied to the transition from ionised to neutral regions that calculations 
  based on the optical depth parametrisation 
  were unable to capture. 
We also present 
  analytical approximations of the calculated 
  spectral features of the \HII bubbles. 
The multiple-ray calculations 
  show that the apparent shape 
  of a \HII bubble (of $d=5$ Mpc at $z=8$),  
  because of the finite speed of light, differs 
  depending on whether the bubble's ionisation front is stationary or expanding. 
Our study shows the necessity 
  of properly accounting for 
  the effects of line-continuum interaction, line broadening 
  and 
  cosmological expansion  
  to correctly predict the EoR 21-cm signals. 
%< 250 words 

\end{abstract}


%%%%%%%%%%%%%%%%%%%%%%%%%%%%%%%%%%%%%%%%%%%%%%%%%%

%%%%%%%%%%%%%%%%% BODY OF PAPER %%%%%%%%%%%%%%%%%% 

%%%%%%%%%%%%%%%%%%%%%%%%%%%%%%%%%%%%%%%%%%%%%%%%%%
%%%%%%%%%%%%%%%%%%%%%%%%%%%%%%%%%%%%%%%%%%%%%%%%%%
% =====================
\section{Introduction} 
\label{sec:introduction}
% =====================
   
At about 0.3 Myr after the Big Bang, electrons and protons began to combine to form neutral hydrogen (\HI) atoms~\citep{Planck2014AA}. This allows the photons to decouple from the charged particles and form the relic radiations which then evolved to become the Cosmological Microwave Background (CMB) observed today. The Universe entered a dark age before the first luminous objects began to form. The UV photons from the first stars and the X-rays from the first active galactic nuclei (AGN) are capable of ionising \HI gas, carving out ionised hydrogen (\HII)  cavities~\citep[see e.g.][]{Loeb2001ARAA}. As these ionised cavities expand~\citep[][]{Furlanetto2004ApJ,Furlanetto2006MNRAS} and merge~\citep[][]{Iliev2006MNRAS,Robertson2010Nat}, the Universe eventually became almost completely reionised, except in some dense self-shielded regions~\citep[see e.g.][]{Chardin2018MNRAS}. This Epoch of Reionisation (EoR) is believed to start at $z\sim$14 and complete at $z\sim$6 based on observations of high redshift luminous quasars~\citep[][]{Fan2006ARAA,Robertson2010Nat} and large-scale polarisation of the CMB~\citep{PlanckXLVI2016AA,PlanckXLVII2016AA}.

These observational constraints, however, suffer from various shortcomings, making it difficult for us to obtain a detailed history of reionisation~\citep[see e.g.][]{Natarajan2014PTEP,Mesinger2016Book}. It can be compensated by observing the 21-cm signals produced during the EoR~\citep[][]{Morales2010ARAA,Baek2010AA,Eide2018MNRAS,Pritchard2012RPPhy,Eide2020MNRAS}. The 21-cm signals are generated by the spin-flip of \HI atoms in the ground state, and corresponds to a frequency of $\nu_{21{\rm cm}} = 1.42~{\rm GHz}$ 
\citep{Hellwig1970IEEE, Essen1971Nat}. 
Due to the abundance of \HI gas of the Universe, we will be able to access the full history of the reionisation via the 21-cm signals when we overcome the difficulties in observation and data analysis. Several projects including the Low Frequency Array 
\citep[LOFAR,][]{Zarka2012sf2a,Mertens2020MNRAS},  Murchison Widefield Array~\citep[MWA,][]{Tingay2013PASA,Trott2020MNRAS}
the Hydrogen Epoch of Reionization Array 
\citep[HERA,][]{DeBoer2017PASP,Abdurashidova2022ApJ_HERA} 
and the Square Kilometer Array 
\citep[SKA,][]{Koopmans2015aska} are underway.

Currently, most of the studies focus on simulating luminous sources and their contributions to reionisation~\citep{Santos2010ascl,Mesinger2011MNRAS,Xu2016ApJ,Park2019MNRAS,Mangena2020MNRAS,Gillet2021MNRAS,Doussot2022AA}. The calculation of 21-cm signals are merely analytical post-processing of simulation results based on the optical depth  parametrisation 
\citep{Furlanetto2006PhR,Pritchard2012RPPhy}. 
Adopting this parametrisation implies that 
the observed 21-cm signal at $z = 0$ at a specific frequency $\nu_{0}$ unambiguously reflects the properties of \HI gas at redshift $z=\nu_{\rm 21cm}/\nu_{0}-1$  
along the line-of-sight (LoS). Line broadening and radiative transfer effects are not correctly accounted for in this parametrisation~\citep[see also][]{Chapman2019MNRAS}.

To quantify the 
inaccuracies of the 21 cm signals computed with the optical depth parameterisation, we adopt a covariant radiative transfer formalism which does not require various assumptions which were used in the derivation of the 21-cm optical depth. 
%difference between the optical depth parametrisation and full radiative transfer of 21-cm
We investigate the spectral features imprinted by ionised cavities on the 21-cm spectra at $z=0$. As most simulations study reionisation on scales larger than galaxies ($\sim 1$~kpc) and present the statistical properties of reionisation, we do not use their detailed ionisation structures. Instead, we construct more generic descriptions for evolving ionised cavities in an expanding Universe which is applicable for all length scales and various ionising sources, in Section~\ref{sec:evolving_cavity}. To facilitate the comparison between the covariant method and the optical depth paramertrisation method, 
we use the most simplified fully ionised spherical cavities as input ionisation structure and calculate their imprints on the 21-cm spectra at $z=0$. We use C21LRT (cosmological 21-cm line radiative transfer) code which tracks the 21 cm signal in both frequency and redshift space, and takes full account of cosmological expansion, radiative transfer effects, and possible dynamics and kinetic effects, such as those caused by macroscopic bulk flow and microscopic turbulence~\citep{Chan2023MNRAS}. The C21LRT formulation and the computational set-up of radiative transfer calculaitions are presented in  
Section~\ref{sec:setup}. We then analyse the spectral features of fully ionised cavities in the absorption and emission regimes of 21-cm global signals and discuss the implications in Section~\ref{sec:resultsDiscussion}. Finally, we summarise the findings in Section~\ref{sec:conclusion}.

%%%%%%%%%%%%%%%%%%%%%%%%%%%%%%%%%%%%%%%%%%%%%%%%%%
%%%%%%%%%%%%%%%%%%%%%%%%%%%%%%%%%%%%%%%%%%%%%%%%%%
% =====================
\section{Ionised Cavities and time evolution} 
\label{sec:evolving_cavity}  

The physics of ionisation processes 
  and the dynamic models of 
  ionisation-front expansion 
  can be found in \citet{Axford1961RSPTA} (see also, \citet{Newman1968ApJa,Yorke1986ARAA,Franco2000ApSS}). 
For the purposes of this study, 
  a generic description 
  for the ionisation bubbles 
  and their expansion is sufficient. 
However, we summarise all the physical processes 
  pedagogically for completeness of this paper. 
We then discuss whether the corresponding physics are incorporated in reionisation simulations and the implications for the accuracy of the predicted 21 cm signals.

%%%%%%%%%%%%%%%%%%%%%%%%%%%%%%%%%%%%%%%%%%%%%%%%%%
% --------
\subsection{Model ionised cavities} 
\label{subsec:model_cavity} 
% --------
We only consider H atoms (\HI and \HII) in our cavity models and C21LRT calculations without any other species. In the following texts, we use `\HI zone' and `\HII zone' for regions filled with only  \HI or \HII gas, respectively, `partially ionised zone' for regions with both \HI and \HII gas, and `ionised cavities' refers to more general/realistic ionisation structures.  The \HII zones are assumed to be spherical. 
Although they would not be spherical in reality, 
  adopting this idealised geometry is sufficient  
  for the purpose of this study, 
  that is, to demonstrate the development of the 21-cm line 
  propagating through evolving ionised cavities.  
  which are driven by 
  continuing ionisation together with cosmological expansion.  

For a \HII zone surrounded by an infinitely extended \HI zone, its size increase are  
due to (i) photoionisation, which is caused 
  by the radiation 
  emitted from the sources embedded inside the \HII zone, 
  and (ii) length stretching as a consequence of cosmological expansion.  
We consider first a two-zone model, 
  which has an insignificant 
  thickness of the transitional interface  
  between the \HII zone
  and the \HI zone. 
We next consider a three-zone model. 
It has a \HII zone 
  surrounded by infinitely extended \HI zone, and a
layer of partially ionised gas between the \HI zone and \HII zone.  
This partially ionised zone 
  has sufficient contribution 
  to the 21-cm line opacity.  
In this study, 
  we consider that the temperature within each zone 
  is uniform at any instant. 
Thus, this simplification 
  does not ignore the temperature evolution 
  of the gases in each of the zones. 

\begin{figure}
%\vspace*{0.2cm} 
    \centering 
%\vspace*{7cm}
    \includegraphics[scale=0.185]{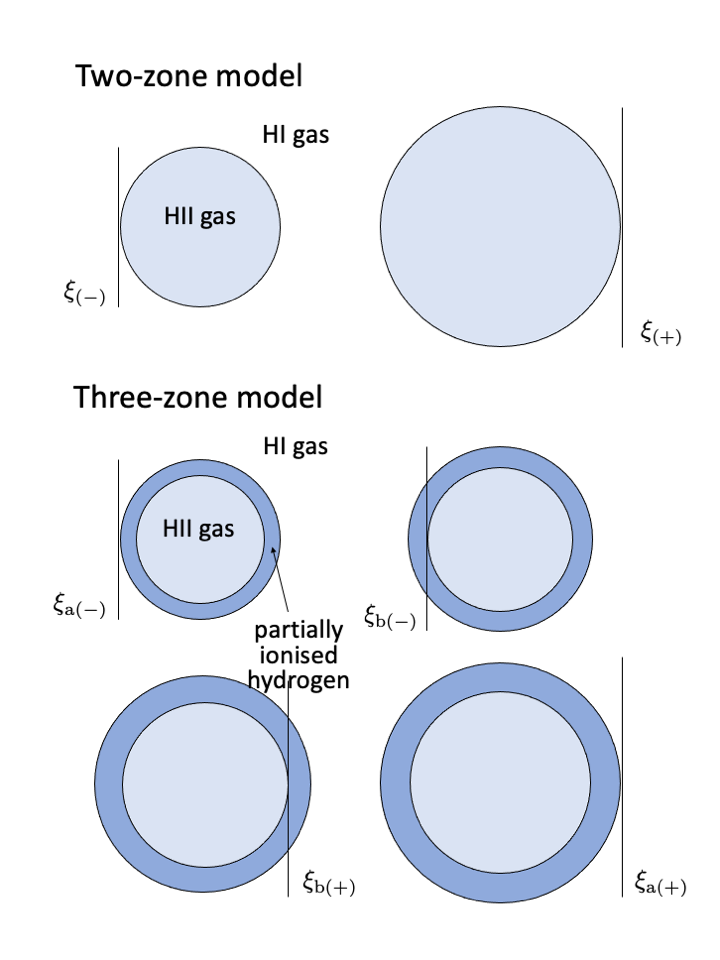}
\vspace*{-0.3cm}  
\caption{A schematic illustration of the two-zone 
  and the three-zone models.  
 The vertical lines are the locations of the light fronts. 
 In the two-zone model (top row), 
   they are are marked by $\xi_{(-)}$ and $\xi_{(+)}$, 
   at the moments when the front reaches and leaves the \HII zone 
   respectively (at $y = 0$). 
 In the three-zone model (middle and bottom rows), 
   they are marked by $\xi_{{\rm a}(-)}$ and $\xi_{{\rm b}(-)}$,  
    $\xi_{{\rm b}(+)}$ and $\xi_{{\rm a}(+)}$, 
   at the moments when the front reaches  
    the outer boundary of the partially ionised zone, 
    finishes the first passage through the partially ionised zone,  
    starts the second passage through the partially ionised zone, 
    and leaves the partially ionised zone, respectively 
    (at $y=0$).   
} 
\label{fig:zone-models}
\end{figure}

For both the two-zone and three-zone models, 
 all zones expand only radially. 
The thickness of the partially ionised zone 
  also increases radially in the three-zone model.   
To avoid using excessive number of parameters, 
  which may cause unnecessary complications 
  that can mask the physics 
  and the 21-cm line structure, 
  we consider that the partially ionised zone 
  is also uniform at any instant. 
A schematic illustration of 
the two-zone and the three-zone models 
is shown in Fig.~\ref{fig:zone-models}.

% Figure 1 
%%%%%%%%%%%%%%%%%%%%%%%%%%%%%%%%%%%%%%%%%%%%%%%%%%
%%%%%%%%%%%%%%%%%%%%%%%%%%%%%%%%%%%%%%%%%%%%%%%%%%  
%%%%%%%%%%%%%%%%%%%%%%%%%%%%%%%%%%%%%%%%%%%%%%%%%%
%%%%%%%%%%%%%%%%%%%%%%%%%%%%%%%%%%%%%%%%%%%%%%%%%%    

%%%%%%%%%%%%%%%%%%%%%%%%%%%%%%%%%%%%%%%%%%%%%%%%%%
% --------
\subsection{Ionisation induced expansion} 
\label{subsec:ionisation_expansion}
% --------
We first consider the situation 
  that the time for light traversing the \HII zone 
  is negligible compared to the Hubble time.  
In this situation, cosmological effects can be ignored, 
  and the radiative transfer of the 21-cm line 
  is subject 
  only to the dynamical evolution of the \HII zone, 
  which is defined by the propagation of the ionisation front.  
Suppose that the \HII zone has a radius $r_0$ initially. 
It expands radially at a speed $v$. 
The radius of the \HII zone at time $t$ is then 
  $r(t) = r_0 + \beta\;\! \xi(t)$ 
  where $\beta = v/c$,  $\xi(t) = c\;\!\!t$, 
  and $c$ is the speed of light. 
Now consider a plane light front propagating   
  in the $x$-direction (in the Cartesian coordinates)  
  from an initial reference location, $x_0$. 
We may set $x_0 = - r_0$.  
Then the light front will reach $x(t) = - r_0 + \xi(t)$ at time $t$.  
It follows that, on the $xy$-plane,  
  the light front will reach the \HII zone boundary at the location $(x,y)$, 
   given by  
\begin{align}  
\begin{cases} 
\hspace*{0.1cm}  
x & = -r_0 + \xi \ ;   \\ 
\hspace*{0.1cm}  
y & = \pm \Big\{ \;\! (1+\beta)\;\!\xi \  \big[\;\! 2r_0 - (1- \beta)\;\!\xi \;\! \big] 
  \;\! {\Big\}}^{1/2} \  ,  
\end{cases}
\label{eq:xy}
\end{align} 
Clearly, at $t =0$, $(x,y) = (-r_0, 0)$. 
Also, we have  
\begin{align} 
  \xi_{(\pm)}\;\!\big\vert_y & = 
  \frac{r_0}{(1-\beta)} 
  \left\{ \;\! 
   1 \pm \left[ 1 - 
   \left( \frac{1-\beta}{1+\beta} \right) 
   \left( \frac{y}{r_0} \right)^2 \right]^{1/2}
 \;\!  \right\} \ , 
\label{eq:xi_2values} 
\end{align}  
implying that the path 
that the light front traverses 
  through the \HII zone is 
\begin{align}  
  \Delta \xi \ \big\vert_y
  & = \xi_{(+)} - \xi_{(-)} = 
   \frac{2\;\! r_0}{(1-\beta)} \;  
   \left[ \;\! 1 - 
   \left( \frac{1-\beta}{1+\beta} \right) 
   \left( \frac{y}{r_0} \right)^2 \;\!
   \right]^{1/2}  \ . 
\label{eq:delta_xi_y}
\end{align} 
The time that the light front meets the \HII zone boundary 
  for a given $y$ therefore has two values, 
  and they are 
  $t_{(\pm)}\vert_y = c^{-1}\xi_{(\pm)}\vert_y$.  
For a non-expanding \HII zone ($\beta = 0$), 
\begin{align}  
  \Delta \xi \  \big\vert_y  & = 
  2\;\! r_0  \ 
   \left[\;\! 1 - \left( \frac{y}{r_0} \right)^2\;\! \right]^{1/2}  \ . 
\label{eq:delta_xi_0}
\end{align}  
This means that expansion leads to LoS length distortion, in terms of a length ratio ${\cal R}$:
\begin{align}  
 {\cal R}  = \left(\frac{1}{1-\beta}\right) 
 \left[   \frac{1- \left( \frac{1-\beta}{1+\beta}\right) 
 \ \left( \frac{y}{r_0} \right)^2}
     {1- \left( \frac{y}{r_0} \right)^2   } \right]^{1/2} \ .
\end{align} 
Note that $\Delta \xi \ \big\vert_y = 0$ occurs at 
\begin{align}
y_{*} & = \pm\;\! r_0 \ \left(\frac{1+\beta}{1- \beta} \right)^{1/2} \ , 
\label{eq:y_values}
\end{align}
 instead of at $y = \pm\;\! r_0$, 
 the values expected for a non-expanding \HII zone 
 -- it has grown bigger 
 since the light front first intersects the \HII zone boundary.  
Note also that the light path $\Delta \xi$ 
  and hence the light traversing time $\Delta t$ diverge  
  when $\beta \rightarrow 1$ 
  (see Equation~\ref{eq:delta_xi_y}).   
This corresponds to the situation where 
  the \HII zone is expanding so rapidly, in the speed of light, 
  that light propagating from the boundary of one side of the \HII zone 
  will never reach the boundary of the other side of the \HII zone, as seen by a distant observer.  

In the two-zone model, the opacity of the \HII zone 
  is provided by the free-electron through Compton scattering. 
The scattering optical depth of radiation 
transported across the  \HII zone at $y$ is  
\begin{align} 
  \tau_{\rm sc} \ \big\vert_y  
  & = (n_{\rm e}\;\! \sigma_{\rm Th})\  \Delta \xi\;\! \big\vert_y  \ ,  
\label{eq:tau-expand}
\end{align}
where $n_{\rm e}$ is the electron number density 
and $\sigma_{\rm Th}$ is the Thompson cross section. 
This can be generalised for ionised cavities with any arbitrary opacity $\kappa_\nu$, 
  at frequency $\nu$, 
  which gives a specific optical depth  
\begin{align}
  \tau_\nu \;\! \big\vert_y  
  & = \kappa_\nu\  \Delta \xi\;\! \big\vert_y =  
   \frac{2\;\!  \kappa_\nu r_0} {(1-\beta)} \;  
   \left[ \;\! 1 - 
   \left( \frac{1-\beta}{1+\beta} \right) 
   \left( \frac{y}{r_0} \right)^2 \;\!
   \right]^{1/2}    \ ,  
\label{eq:tau_nu-expand}
\end{align} 
 which is in contrast to the specific optical depth expected 
 for a non-expanding cavity:  
\begin{align}
  \tau'_\nu \;\! \big\vert_y  & = 
  {2\;\!  \kappa_\nu r_0}  \;  
   \left[ \;\! 1 - 
   \left( \frac{y}{r_0} \right)^2 \;\!
   \right]^{1/2}    \ .  
\label{eq:tau_nu-noexpand}
\end{align} 

The remaining question is now how fast the \HII zone would expand. 
We may obtain a rough estimate for the expansion speed $v$ 
 from the following consideration.  
The boundary of the \HII zone advances  
 only when the amount of ionising radiation 
 is sufficient to convert \HI atoms to ions. 
That is, 
\begin{align} 
 \beta & = \frac{v}{c} 
 \approx \left( 1 - \frac{\Upsilon n_{\rm HI}(r)}{n_\gamma(r)}
 \right)   \ , 
\label{eq:saturation}
\end{align}
 where $n_\gamma$ is the number density of ionising photons, 
 and $n_{\rm HI}$ is the number density of \HI in the surrounding.  
The efficiency of converting \HI atoms into ions 
  may parameterised by a variable $\Upsilon$, 
  which is determined by the local atomic and thermodynamic properties 
  (i.e. not the global geometry of the system) 
  and the ratio $(n_{\rm HI}/n_\gamma)$.   
The expression in Eqn.~(\ref{eq:saturation}) 
  is essentially a measure of how saturated the ionisation process is. 

The number density of ionising photon $n_\gamma (r)$ 
  attenuates with $r$ 
  due to the dilution of radiation over distance 
  and the consumption of the ionising photons to facilitate 
  the expansion. 
Thus, the expansion of the \HII zone even slows down 
  as ionisation approaches saturation, where 
  processes, such as recombination, 
  counteract ionisation 
  when the photon supply becomes insufficient. 
Nonetheless, in the very initial stage  
  when there is a sudden burst of ionising sources, 
  the supply of $n_\gamma$ is abundant, 
  and hence the ionisation is far from saturation. 
For $n_\gamma \gg \Upsilon n_{\rm HI}$, 
 the expansion velocity $v$ would be expected to almost reach the speed of light, 
 i.e. $\beta \lesssim 1$. 
For modest unsaturated ionisation, 
  which gives rise a (constant) normalised expansion velocity $\beta =0.5$, gives 
  $|y_*| = \sqrt{3}\ r_0$, and 
   $\tau_\nu = 4 \;\! \kappa_\nu r_0 = 2 \tau'_\nu$ at $y = 0$ 
   (i.e. radiative transfer along the symmetry axis of the \HII zone).   
Increasing the expansion velocity to $\beta = 0.75$ 
  gives $|y_*| = \sqrt{7}\ r_0$, and 
  $\tau_\nu = 8\;\! \kappa_\nu r_0 = 4 \tau'_\nu$ at $y = 0$.  
Here, it shows that ignoring the expansion of \HII zone 
  could lead to incorrect inferences 
  for its size and the optical depth. 
 
The situation is slightly more complicated, 
  if there is a partially ionised zone enveloping  
  the \HII zone. 
The partially ionised zone 
  determines the radiative transfer process 
  together with the \HII zone. 
The temperature in the partially ionised zone  
 could be an important variable, 
 especially in the context of radiative transfer of the 21-cm line,  
 as its relative contrast with the temperatures 
 of the background \HI gas  
 and the temperature of the \HII gas 
 would determine whether a line 
 would appear as emission or as absorption 
 relative to the continuum radiation. 
Apart from the microscopic aspects, 
  such as radiation processes and radiative transfer, 
  there are subtle issues in the macropscopic perspectives. 
The inner and outer boundary of the partially ionised zone   
  could expand asynchronously at different speeds.  
There are also additional issues, 
  which is generally insignificant for the development of the \HII zone 
  but might need to be take into account for the development of the partially ionised zone. 
This is due to the fact that the boundary of the \HII zone
is an ionisation front, not a hydrodynamics shock front, 
which needs to satisfy certain shock-jump condition, 
  while the interface between the partially ionised zone 
  and the \HI zone  
  could be determined by hydrodynamics and also thermodynamic processes. 
Nonetheless, we leave the investigation of these subtle yet important physics  
  in a future paper. 
In this section, we employ a simple parametric model 
  to illustrate how the presence of an additional partially ionised zone 
  would alter the radiative transfer process.   
 
Assume that the \HII zone and the partially ionised 
zone expand in uniform normalised speeds:  
the outer boundary of the partially ionised zone has an initial radius $r_{\rm a0}$ and expand with $\beta_{\rm a}$. 
The boundary of the \HII zone has an initial radius $r_{\rm b0}$ and expands with and $\beta_{\rm b}$
  (with $\beta_{\rm b} \leq \beta_{\rm a}$, 
   which avoids the unphysical situation where 
   the expansion of the \HII zone overruns 
    the expansion of the partially ionised zone).  
At $t =0$,   
  when the light front, propagating in the $x$-direction 
  (as in the two-zone model) reaches the outer boundary of the partially ionised zone, 
  which is located at $x = - r_{\rm a0}$. 
The results obtained above for the two-zone model 
  can be adopted using the substitutions   
  $r_0 \rightarrow r_{\rm a0}$ and $\beta \rightarrow \beta_{\rm a}$. 
The advancing of the light front is now 
  $x(t) = - r_{\rm a0} + \xi(t)$, 
  and the radius of the outer boundary of the partially ionised zone
  is  $r(t) = r_{\rm a0} + \beta_{\rm a}\xi(t)$. 
Thus,   
\begin{align}  
\begin{cases} 
\hspace*{0.1cm}  
x & = -r_{\rm a0} + \xi \ ;   \\ 
\hspace*{0.1cm}  
y & = \pm \Big\{ \;\! (1+\beta_{\rm a})\;\!\xi \;\!
 \big[\;\! 2r_{\rm a0} - (1- \beta_{\rm a})\;\!\xi \;\! \big] 
  \;\! {\Big\}}^{1/2} \  .  
\end{cases}
\label{eq:xy-outer}
\end{align}  
Also,  we have 
\begin{align} 
  \xi_{\rm a{(\pm)}}\;\!\big\vert_y & = 
  \frac{r_{\rm a0}}{(1-\beta_{\rm a})} 
  \left\{ \;\! 
   1 \pm \left[ 1 - 
   \left( \frac{1-\beta_{\rm a}}{1+\beta_{\rm a}} \right) 
   \left( \frac{y}{r_{\rm a0}} \right)^2 \right]^{1/2}
 \;\!  \right\} \ .  
\label{eq:xi_2values-outer} 
\end{align}  
The corresponding maximum size of the partially ionised zone perceived by the distant observer is  
\begin{align}
y_{*{\rm a}} & = \pm\;\! r_{\rm a0} \ 
 \left(\frac{1+\beta_{\rm a}}{1- \beta_{\rm a}} \right)^{1/2} \ , 
\label{eq:y_values_outer}
\end{align}  

For the inner boundary of  
  the partially ionised zone, 
  some additional transformations for the corresponding variables 
  are required, 
  before applying the results obtained for the two-zone model. 
First,  
  the light front reaches this boundary at a time $\tilde t$, 
  at a location between $x = -r_{\rm a0}$ and $x = -r_{\rm b0}$. 
Denote this location as $x = - {\tilde r}_{\rm b0}$. 
We may then determine ${\tilde r}_{\rm b0}$, 
 $\tilde t$, and hence $\tilde \xi (= c {\tilde t})$ 
 by setting $-r_{\rm a0} + {\tilde \xi}  =  - r_{\rm b0} - \beta_{\rm b}{\tilde \xi}$.  
This gives  
\begin{align}  
  {\tilde \xi} & = c\;\! {\tilde t} 
  = \frac{r_{\rm a0} - r_{\rm b0}}{(1+ \beta_{\rm b})} 
  = r_{\rm a0} \left(\frac{1-\varpi}{1+\beta_{\rm b}} \right) \ ,  
\end{align} 
and 
\begin{align}
{\tilde r}_{\rm b0} & = 
 \frac{\beta_b r_{\rm a0}+r_{\rm b0} }{(1+\beta_b)}  
  = r_{\rm a0}  \left(\frac{\varpi+\beta_{\rm b}}{1+\beta_{\rm b}} \right)\ , 
\end{align}  
where $\varpi = r_{\rm b0}/r_{\rm a0} \leq 1$. 
With the substitutions of $r_{\rm a0} \rightarrow {\tilde r}_{\rm b0}$, 
 $\beta_a \rightarrow \beta_{b}$   
  and $\xi \rightarrow (\xi - {\tilde \xi})$  
  in Eqns.~(\ref{eq:xy-outer}), (\ref{eq:xi_2values-outer}) and (\ref{eq:y_values_outer}), 
  we obtain  
\begin{align}  
\begin{cases} 
\hspace*{0.1cm}  
x & = -r_{\rm a0} + \xi = -{\tilde r}_{\rm b0}+(\xi - {\tilde \xi})\ ;   \\ 
\hspace*{0.1cm}  
y & = \pm \Big\{ \;\! (1+\beta_{\rm b})\;\!\left(\xi - {\tilde \xi}\right)   
 \big[\;\! 2{\tilde r}_{\rm b0} - (1- \beta_{\rm b})
 \;\!\left(\xi-{\tilde \xi}\right) 
  \;\! \big] 
  \;\! {\Big\}}^{1/2} \  ,   
\end{cases}
\label{eq:xy-inner}
\end{align} 
 for the interception of the light front 
 with the boundary of the \HII zone. 
It follows that  
\begin{align} 
  \xi_{\rm b{(\pm)}}\;\!\big\vert_y - {\tilde \xi} & = 
  \frac{{\tilde r}_{\rm b0}}{(1-\beta_{\rm b})} 
  \left\{ \;\! 
   1 \pm \left[ 1 - 
   \left( \frac{1-\beta_{\rm b}}{1+\beta_{\rm b}} \right) 
   \left( \frac{y}{{\tilde r}_{\rm b0}} \right)^2 \right]^{1/2}
 \;\!  \right\} \ , 
\label{eq:xi_2values-inner-1} 
 \end{align} 
  implying that 
\begin{align}
 \xi_{\rm b{(\pm)}}\;\!\big\vert_y 
  & =  r_{\rm a0} \left\{ 
  \left(\frac{1-\varpi}{1+\beta_{\rm b}} \right) + 
  \left(\frac{\varpi+\beta_{\rm b}}{1-{\beta_{\rm b}}^2} \right)\;\! 
  \big[ \;\! 
   1 \pm f(y; \beta_{\rm b}, \varpi, r_{\rm a0}) 
 \;\!  \big] \right\}  \ , 
\label{eq:xi_2values-inner-2} 
\end{align}
 where 
\begin{align}   
 f(y; \beta_{\rm b}, \varpi, r_{\rm a0}) & = 
   \left[ 1 - 
  \left( 
   \frac{1-{\beta_{\rm b}}^2}{\left(\varpi+\beta_{\rm b}\right)^2} 
   \right) 
   \left( \frac{y}{ r_{\rm a0}}  \right)^2 \right]^{1/2}\ \leq 1 \ ,    
\label{eq:xi_2values-inner-ex} 
\end{align} 
  and is always positive.   
Also, 
\begin{align}
y_{*{\rm b}} & = \pm\;\! {\tilde r}_{\rm b0} \ 
 \left(\frac{1+\beta_{\rm b}}{1- \beta_{\rm b}} \right)^{1/2} 
 = \pm\;\! r_{\rm a0} \ \left( 
   \frac{\left(\varpi+\beta_{\rm b}\right)^2}{1-{\beta_{\rm b}}^2} 
   \right)^{1/2}  \ .  
\label{eq:y_values_inner}
\end{align}

In the context of radiative transfer (and ray-tracing), 
  a fraction of the radiation 
  (those at $|y_{*{\rm a}}| \geq |y| > |y_{*{\rm b}}|$) 
    will pass through the partially ionised zone, 
  and another fraction of the radiation 
  (those at $|y_{*\rm b}| \geq |y|$) 
  will pass through the partially ionised zone
    before entering the \HII zone 
    and then pass through the partially ionised zone later afterwards. 
For the rays at $|y_{*{\rm a}}| \geq |y| > |y_{*{\rm b}}|$, 
  the specific optical depth, at frequency $\nu$, across the \HII and partially ionised zones is 
 \begin{align}
  \tau_\nu \;\! \big\vert_y  
  &  =  
   \frac{2\;\!  \kappa_{\nu {\rm a}} r_{\rm a0}} 
     {(1-\beta_{\rm a})} \;  
   \left[ \;\! 1 - 
   \left( \frac{1-\beta_{\rm a}}{1+\beta_{\rm a}} \right) 
   \left( \frac{y}{r_{\rm a0}} \right)^2 \;\!
   \right]^{1/2}    \ ,    
\label{eq:tau_nu-3zone_a}
\end{align}  
 where $\kappa_{\nu {\rm a}}$  
  is the specific opacity of the gas in the partially ionised zone.  
For rays with $|y_{*\rm b}| \geq |y|$, 
  the specific optical depth is the sum of three components: 
  two from the partially ionised zone and one form the \HII zone. 
Along the ray, they are the segments of the light path 
 specified by $(\zeta_{\rm a(-)},\zeta_{\rm b(-)})$  
 for the first passage in the partially ionised zone, 
 $(\zeta_{\rm b(-)},\zeta_{\rm b(+)})$   
 for the passage in the \HII zone,  
 and $(\zeta_{\rm b(+)},\zeta_{\rm a(+)})$   
 for the second passage in the partially ionised zone. 
We denote the corresponding specific optical depth for these three passages 
  as $\tau^{[{\rm i}]}_\nu$, $\tau^{[{\rm j}]}_\nu$ 
  and $\tau^{[{\rm k}]}_\nu$  
  respectively, 
  and the total specific optical depth is the sum of them, i.e. 
  $\tau_\nu = 
  \tau^{[{\rm i}]}_\nu + \tau^{[{\rm j}]}_\nu + \tau^{[{\rm k}]}_\nu$.   

Using similar procedures 
  as in the derivation of Eqn.~(\ref{eq:tau_nu-3zone_a}), 
  we obtain 
\begin{align} 
  \tau^{[{\rm j}]}_{\nu}\;\!\big\vert_y 
   & = \;\! 2\;\! \kappa_{\nu {\rm b}} r_{\rm a0}
     \left(  \frac{\varpi + \beta_{\rm b}}{1-{\beta_{\rm b}}^2}   \right) 
   \left[ \;\! 1 - 
   \left( 
   \frac{1-{\beta_{\rm b}}^2}{\left(\varpi+\beta_{\rm b}\right)^2} 
   \right) 
   \left( \frac{y}{r_{\rm a0}} \right)^2 \;\!
   \right]^{1/2}    \nonumber \\ 
  & = \;\! 2\;\! \kappa_{\nu {\rm b}} r_{\rm a0}
     \left(  \frac{\varpi + \beta_{\rm b}}{1-{\beta_{\rm b}}^2}   \right) 
      \ f(y; \beta_{\rm b}, \varpi, r_{\rm a0})
      \label{eq:op-j}
%\label{eq:tau_nu-3zone_bj}
\end{align}
    for the \HII zone\footnote{We 
    can recover Eqn.~(\ref{eq:tau_nu-3zone_a}) 
    by setting $\varpi \rightarrow 1$, 
    $\beta_{\rm b} \rightarrow \beta_{\rm a}$ and 
    $\kappa_{\nu {\rm b}} \rightarrow \kappa_{\nu {\rm a}}$ 
    in Eqn.~(\ref{eq:op-j}).}.  
To derive the specific optical depths 
  of the two passages through the partially ionised zone, 
  we may consider the expression  
\begin{align} 
  \xi_{{\rm a}(\pm)}\;\!\big\vert_y & = 
    \frac{r_{\rm a0}}{(1-\beta_{\rm a})} 
      \left\{ 1 \pm f(y;\beta_{\rm a},1,r_{\rm a0}) \right\} \ . 
\label{eq:xi_2values-outer-3}
\end{align} 
 to simplify the algebraic steps. 
The specific optical depth 
   corresponding to the first passage is  
\begin{align} 
\tau^{[{\rm i}]}_\nu \;\! \big\vert_y & =  
  \kappa_{\nu{\rm a}}\;\! \big[\;\!\xi_{{\rm b}(-)} - \xi_{{\rm a}(-)} 
  \;\! \big] \nonumber \\  
  & = \kappa_{\nu{\rm a}}r_{\rm a0} 
  \Bigg\{  
  \left(\frac{1-\varpi}{1+\beta_{\rm b}} \right) + 
  \left(\frac{\varpi+\beta_{\rm b}}{1-{\beta_{\rm b}}^2} \right)\;\! 
  \big[ \;\! 
   1 - f(y; \beta_{\rm b}, \varpi, r_{\rm a0}) 
 \;\!  \big]  \nonumber \\ 
 & \hspace*{1.5cm} - 
  \left(\frac{1}{1-\beta_{\rm a}} \right)\;\! 
  \big[ \;\! 
   1 - f(y; \beta_{\rm a}, 1, r_{\rm a0}) 
 \;\!  \big] \Bigg\}  \ ;    
\label{eq:op-i}
\end{align} 
the specific optical depth 
   corresponding to the second passage is   
\begin{align} 
\tau^{[{\rm k}]}_\nu \;\! \big\vert_y & =  
  \kappa_{\nu{\rm a}}\;\! \big[\;\!\xi_{{\rm a}(+)} - \xi_{{\rm b}(+)} 
  \;\! \big] \nonumber \\  
  & = \kappa_{\nu{\rm a}}r_{\rm a0} 
  \Bigg\{  \left(\frac{1}{1-\beta_{\rm a}} \right)\;\! 
  \big[ \;\! 
   1 + f(y; \beta_{\rm a}, 1, r_{\rm a0})  \;\!  \big] 
  - \left(\frac{1-\varpi}{1+\beta_{\rm b}} \right) 
   \nonumber \\ 
  & \hspace*{1.5cm}   - 
  \left(\frac{\varpi+\beta_{\rm b}}{1-{\beta_{\rm b}}^2} \right)\;\! 
  \big[ \;\! 
   1 + f(y; \beta_{\rm b}, \varpi, r_{\rm a0})  
 \;\!  \big] \Bigg\}  \ . 
\label{eq:op-k}
\end{align}  
With non-zero $\kappa_{\nu{\rm a}}$ and $\kappa_{\nu{\rm b}}$, 
 summing the specific optical depths 
  in Eqns.~(\ref{eq:op-j}), (\ref{eq:op-i}) and (\ref{eq:op-k}) yields 
\begin{align} 
 \tau_\nu \;\! \big\vert_y  
  & = 2\kappa_{\nu{\rm a}}r_{\rm a0} 
  \Bigg\{ 
  \left( \frac{1}{1-\beta_{\rm a}} \right) 
   f(y;\beta_{\rm a}, 1, r_{\rm a0})  \nonumber \\ 
   & \hspace*{1.2cm}
 + \left(\frac{\kappa_{\nu{\rm b}}}{\kappa_{\nu{\rm a}}} - 1 \right) 
  \left( \frac{\varpi +\beta_{\rm b}}{1- {\beta_{\rm b}}^2} \right) 
   f(y;\beta_{\rm b}, \varpi, r_{\rm a0}) \;\! 
  \Bigg\} \  
\label{eq:tau_nu_final}
\end{align} 
 for $|y_{*{\rm b}}| > |y|$.  
Note that the expression for the specific optical depth above 
  becomes the same as that in Eqn.~(\ref{eq:tau_nu-3zone_a}), 
  by equating $\kappa_{\nu {\rm b}}$ and $\kappa_{\nu {\rm a}}$, 
   regardless of what values $\beta_{\rm b}$ takes. 
However, if we set $\varpi \rightarrow 1$ 
  and $\beta_{\rm b} \rightarrow \beta_{\rm a}$ 
  in Eqn.~(\ref{eq:tau_nu_final}), 
  then Eqn.~(\ref{eq:tau_nu-3zone_a})
  becomes a special case of it  
  where $\kappa_{\nu{\rm b}}$ equals $\kappa_{\nu{\rm a}}$.

The structures of 
  real ionised cavities are expected to be more complex than 
  that of the two-zone and three-zone model bubbles. 
Ionised cavities produced by quasars and  galaxies have been studied in detail with analytical~\citep[e.g.\,][]{Shapiro1987ApJ, Wyithe2004ApJ}, semi-numerical simulations~\citep[e.g.\,][]{Geil2008MNRAS,Mesinger2011MNRAS} and hydrodynamical simulations~\citep[e.g.\,][]{Thomas2009MNRAS, Geil2017MNRAS, 
Hutter2021MNRAS,   
Kanna2022MNRAS_THESAN}. While these studies focused on the global progression of reionisation, some also targeted individual bubbles
~\citep[e.g.\,][]{Ghara2017MNRAS}. The statistics properties of the cavities were also  investigated \citep[e.g.][]{Zahn2007ApJ, Shin2008ApJ, Lin2016MNRAS, Munoz2022MNRAS, Schaeffer2023MNRAS, Lu2024MNRAS}. The 21-cm signal associated with complex ionisation structures are not ideal for comparing our covariant formulation with the optical depth parametersation. Hence we adopt the simplified the two-zone and three-zone models.

%%%%%%%%%%%%%%%%%%%%%%%%%%%%%%%%%%%%%%%%%%%%%%%%%%
% --------
\subsection{Cosmological expansion} 
\label{subsec:cosmological expansion}  
% --------
Cosmological expansion affects the observational properties 
  different to the local expansions,  
  such as the geometrical expansion 
  caused by the advance of an ionisation front. 
When the Universe expands, 
  an ionised cavity expands accordingly, 
  even when the ionisation front is stationary 
  in the local reference frame. 
Cosmological expansion 
  also alters the thermodynamics and hydrodynamic properties of the ionised cavities.  
It sets a new balance 
  between the radiative processes, 
  hence modifying 
  the radiation and the observational characteristics 
  of the ionised cavity.
  
Two effects are the most noticeable among the others. 
(i) The size of the cavity  
    perceived by an observer at a lower redshift 
    is larger than 
    the size of the cavity 
    measured in its local reference frame,     
    where the radiative and hydrodynamic processes operates. 
(ii) The wavelength of the radiation 
   is stretched as it  propagates across the cavity. 
A 21-cm line originated from the far side of the cavity 
     will have a wavelength longer than 21~cm 
     when it reaches another side of the cavity.   
This change in the waveleghth 
  will alter the transport of the radiation. 
For the 21-cm line, 
   the interaction with its neighbouring continuum 
   becomes prominent, 
   while the resonance absorption can become insignificant.   
This has not taken account of further complications 
   by effects associated with 
   thermal and hydrodynamics evolution 
   of the cavity and of the medium 
   surrounding the cavity.  

\begin{figure}
%\vspace*{0.2cm} 
    \centering 
%\vspace*{7cm}
    \includegraphics[scale=0.42]{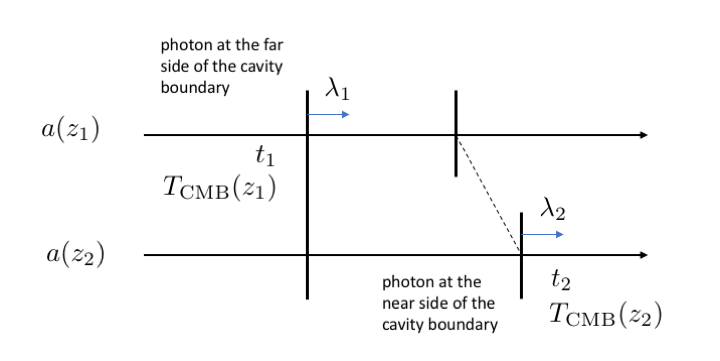}
\vspace*{-0.3cm}  
\caption{An illustration showing the 
 changes in the properties of the radiation, 
 e.g. its wavelength $\lambda$,  
 and its environment, e.g. the CMB temperature $T_{\rm CMB}$, 
 along the LoS  
 between the moment that 
 it started propagating from the far side of the cavity boundary 
 (with variables denoted by the subscript ``1'')
 and the moment that it reached the near side of the cavity boundary 
  (with variables denoted by the subscript ``2''), 
  as perceived by a distant observer at present. 
A useful reference for the changes 
  is the cosmological redshift $z$, 
  which has a 1-1 correspondence to a cosmological time $t$.  
The rate of change in the scale factor $a$ 
  in the FLRW metric  
  is a measure of the cosmological expansion, 
  which may be parameterised by $z$ or $t$. 
The same LoS is drawn twice in the figure  
 to illustrate separately the conditions 
 when the radiation started and finished its propagation 
 across the cavity. 
} 
\label{fig:cos-expansion}
\end{figure}

Fig.~\ref{fig:cos-expansion} shows an illustration to elaborate 
  these effects. 
A photon of a wavelength $\lambda_1$ 
  starts its journey from the far side of a cavity boundary,  
  at redshift $z_1$ (corresponding to a cosmological time $t_1$) 
  and arrives at the near side of the cavity boundary, 
  at a lower redshift $z_2$ (corresponding to the a cosmological time $t_2$). 
When the photon reached the other side of the cavity boundary 
  its wavelength became $\lambda_2$. 
The cavity was bathed in the CMB
  of temperature $T_{\rm CMB}(z_1)$ 
  at the moment the photon started its journey,  
  but the temperature of the ambient CMB 
  had dropped to a lower temperature $T_{\rm CMB}(z_2)$ 
  by the time the photon arrived at the other side of the cavity.

% Figure 2 
%%%%%%%%%%%%%%%%%%%%%%%%%%%%%%%%%%%%%%%%%%%%%%%%%%
%%%%%%%%%%%%%%%%%%%%%%%%%%%%%%%%%%%%%%%%%%%%%%%%%%  
%%%%%%%%%%%%%%%%%%%%%%%%%%%%%%%%%%%%%%%%%%%%%%%%%%
%%%%%%%%%%%%%%%%%%%%%%%%%%%%%%%%%%%%%%%%%%%%%%%%%%    

%%%%%%%%%%%%%%%%%%%%%%%%%%%%%%%%%%%%%%%%%%%%%%%%%
%%%%%%%%%%%%%%%%%%%%%%%%%%%%%%%%%%%%%%%%%%%%%%%%%%  
\begin{figure}
\vspace*{-0.6cm} 
    \centering 
    \includegraphics[scale=0.3]{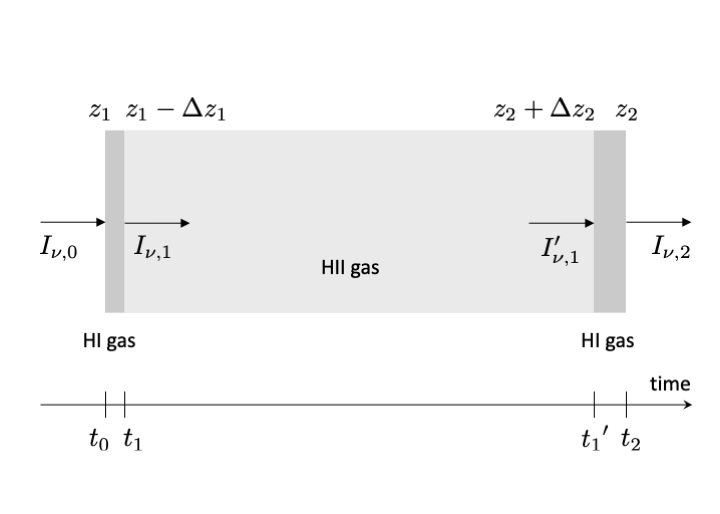}
\vspace*{-0.5cm}  
\caption{A schematic illustration of the three-slab model 
  to show radiative transfer effects 
  on line and continuum of radiation through 
  an \HII slab sandwiched by two 
  initially identical thin \HI slabs 
  in an expanding universe.  
When the radiation enters the first \HI slab, 
  its specific intensity is $I_{\nu,0}$; 
  when the radiation exits, it is  $I_{\nu,1}$. 
 When the radiation enters the second \HI slab, 
  its specific intensity is $I'_{\nu,1}$; 
  when the radiation exits, it is  $I_{\nu,2}$.  
The sequence of redshifts at which the radiation 
  passes the slab boundaries in its propagation 
  is $\{ z_1,\;\! z_1-\Delta z_1,\;\! z_2+\Delta z_2,\;\! z_2 \}$.  
The corresponding sequence of time is 
  $\{ t_0,\;\! t_1,\;\! {t_1}',\;\! t_2  \}$.  
} 
\label{fig:RT_line}
\end{figure} 
%%%%%%%%%%%%%%%%%%%%%%%%%%%%%%%%%%%%%%%%%%%%%%%%%%
%%%%%%%%%%%%%%%%%%%%%%%%%%%%%%%%%%%%%%%%%%%%%%%%%%    

In a Friedmann-Lema\^{i}tre-Robertson-Walker (FLRW) universe with a zero curvature, 
  the expansion of the universe can be parameterised 
  by a scale parameter $a(t)$, 
  which evolves with the cosmological redshift as 
\begin{align} 
\frac{a(t_2)}{a(t_1)} 
  = \frac{1+z_1}{1+z_2} \ . 
\end{align}   
This gives the stretch of the wavelength of radiation  
\begin{align} 
 \frac{\lambda_2}{\lambda_1} & =  \frac{a(t_2)}{a(t_1)}   
   = \frac{1+z_1}{1+z_2} \ , 
\end{align} 
  and the evolution of the CMB temperature  
\begin{align} 
 %\frac{T_2}{T_1} & = 
 \frac{T_{\rm CMB}(z_2)}{T_{\rm CMB}(z_1)} 
 = \frac{1+z_2}{1+z_1}  \ . 
\end{align} 
The number density of particles evolves with the cosmological redshift as  
 \begin{align} 
 \frac{n_2}{n_1} & = \frac{n(z_2)}{n(z_1)} 
  = \left( \frac{1+z_2}{1+z_1}  \right)^3   \ . 
\end{align}

Consider that a photon propagates from a location $r_1$ 
   to a location at $r_2$.  
It starts from $r_1$ at time $t_1$ and reaches $r_2$ at time $t_2$. 
As photons travel along null geodesics,    
 in a flat FLRW universe  
 $c\;\!{\rm d}\;\!t = a(t)\;\! {\rm d}\;\!r$ 
   along the LoS.  
It follows that 
\begin{align}  
\frac{1}{c} \int_{r_1}^{r_2}  {\rm d}r 
  & = \int_{z_1}^{z_2} {\rm d}z \; \! (1+z)  
   \left(\frac{{\rm d}t}{{\rm d}z}\right) \ . 
\label{eq:z2a}
\end{align} 
 Here $z_1$ and $z_2$ 
 are the redshifts, with respect to a present observer located at $r=0$   
 at redshift $z=0$, 
 respectively, and correspond 
 to the time $t_1$ (when the photon starts its journey from $r_1$)
 and $t_2$ (the time when the photon completes its journey reaching $r_2$), 
 with $a(t) = 1$ at $z=0$.  
Hence, $z_2$ satisfies 
\begin{align}  
  \int_{0}^{z_2} {\rm d}z \; \! (1+z)  
   \left(\frac{{\rm d}t}{{\rm d}z}\right) 
   & = \int_{0}^{z_1} {\rm d}z \; \! (1+z)  
   \left(\frac{{\rm d}t}{{\rm d}z}\right)  
   - \left(\frac{r_1-r_2}{c} \right) \ .   
\label{eqn:z2b}
\end{align} 
In the $\Lambda$CDM framework, 
\begin{equation}
\label{eq:dtdz} 
 \frac{{\rm d} t}{{\rm d}z}  
 =  - \Bigg\{ H_0 (1+z) 
  \left[
   \Omega_{{\rm r},0}(1+z)^4 
   + \Omega_{{\rm m},0}(1+z)^3 
   +\Omega_{\Lambda,0}
   \right]^{1/2} 
  \Bigg\}^{-1} \  
\end{equation}  
\citep[see e.g.][]{Peacock1999book}
where $H_0$ is the Hubble parameter at present, and 
  $\Omega_{\rm r,0}$, 
  $\Omega_{\rm m,0}$ 
  and $\Omega_{\Lambda,0}$ 
  are the density parameters for 
  radiation, matter and dark energy, respectively. 
For the spherical \HII zones in the two-zone model 
 without expansion induced by ionisation 
 (Fig.~\ref{fig:zone-models}, see also 
 Sec.~\ref{subsec:ionisation_expansion}), 
 $(r_1 - r_2) = - 2 r_0$ 
 along the symmetry axis. 
With this specified,  
  $z_2$ is readily determined for a given $z_1$, and this can be generalised for the two-zone and three-zone models. 
Assigning the location of the zone boundaries as $r_2$, 
 which specifies the value for $(r_1 - r_2)$. 
The corresponding $z_2$ (and hence $t_2$)   
 can be obtained by a direction integration of 
 Eqn.~(\ref{eqn:z2b}). 
During the Dark Ages and the EoR, 
  the cosmological evolution is matter dominated. 
We may ignore $\Omega_{{\rm r,0}}$ and $\Omega_{\Lambda,0}$, 
  which results in an analytic expression for $z_2$:  
\begin{align}
     z_{2} 
     & = \left( \frac{1}{\sqrt{1+z_{1}}} 
     -\left( \frac{r_1-r_2}{c}\right)
     \frac{H_0{\sqrt{\Omega_{\rm m,0}}}}{2}\right)^{-2} -1 \ .  
 \end{align}

We now use a three-slab model 
  to illustrate the difference between the spectra, 
  consisting of a line component 
  and a continuum component, 
  in the presence and in the absence of cosmological expansion.

Consider a ray of radiation with initial specific intensity $I_{\nu, 0}$, 
  passing through a slab of \HII gas sandwiched 
  by two thermal absorptive slabs of \HI gas,  
  both of co-moving thickness $\delta s$  
  (see Figure~\ref{fig:RT_line}). 
Without losing generality,  
  the total optical depth of the \HII slab is 
  assumed to be negligible, 
  i.e. the slab practically does not contribute 
  to absorption and emission of the radiation. 
The two \HI slabs are transparent to the continuum 
  but have sufficient optical depths 
  at the line centre frequency, $\nu_{\rm 21cm}$, and its neighbouring frequencies. 
The local absorption in the \HI slabs is specified 
 by an absorption coefficient, which takes the form:      
$\kappa_\nu = n_{\rm a}\;\!\sigma_{\rm a}\ \phi(\nu - \nu_{\rm 21cm})$,     
  where $n_{\rm a}$ is the number density of the absorbers, 
    $\sigma_{\rm a}$ is the normalised absorption cross section, 
    $\nu_{\rm 21cm}$ is the centre frequency 
     measured in the local rest frame,   
    and $\phi(\nu - \nu_{\rm 21cm})$ 
    is a normalised line profile function, i.e. 
\begin{align} 
 \int_0^\infty   {\rm d} \nu \ \phi(\nu - \nu_{\rm 21cm}) = 1 \ . 
\end{align} 
A thermal absorptive media also emits 
  and the local emissivity can be derived from 
  the absorption coefficient and the source function, 
  which is the Planck function $B_\nu(T)$ 
  in a thermal medium.     
(Note that, for simplicity, 
  we have omitted the contribution from stimulated emission in this illustration. 
For the radiative transfer calculations 
  of the hyperfine 21-cm \HI 
  line used for diagnosing cosmological reionisation, 
  stimulated emission must be included.)

At low-frequencies (i.e. in the Rayleigh-Jeans regime), 
 the Planck function   
 $B_\nu \rightarrow I{^{\rm RJ}}_{\nu} = b \nu^2 T$, 
 where $b = 2 k_{\rm B} c^{-2}$ and  
 $k_{\rm B}$ is the Boltzmann constant.  
For $\Delta z_1 \ll (1+ z_1)$  and $\Delta z_2 \ll (1+ z_2)$, 
  we may approximate that the emission and absorption 
  are constant with time. 
Suppose also that the background is a thermal black-body of a temperature $T_0$.  
Then, in this situation, 
\begin{align} 
 I_{\nu,1}  & = I_{\nu,0}\;\! 
   \exp(-n_{\rm a,1}\;\!\sigma_{\rm a}\;\!\phi(\nu-\nu_{\rm 21cm}) 
   \;\!\delta s) \nonumber \\ 
  & \hspace*{0.5cm}  
    + B_\nu(T_1)\;\! 
    \Big[\;\! 1- \exp(-n_{\rm a,1}\;\!\sigma_{\rm a}\;\!\phi(\nu-\nu_{\rm 21cm})
    \;\!\delta s )\;\!\Big]  
    \nonumber \\ 
  & = b \nu^2\;\! \Big[\;\!  T_0  \;\! 
   \exp(-n_{\rm a,1}\;\!\sigma_{\rm a}\;\!\phi(\nu-\nu_{\rm 21cm})
   \;\!\delta s) \nonumber \\
   & \hspace*{1.5cm}  
    +  T_1\;\! 
    \left[1- \exp(-n_{\rm a,1}\;\!\sigma_{\rm a}\;\!\phi(\nu-\nu_{\rm 21cm})
    \;\!\delta s )\right]  
    \;\! \Big]  \ , 
\end{align}    
where $n_{\rm a,1}$ is the number density of absorber in \HI slab 1 
  and $T_1$ is the thermal temperature associated 
   with the line-formation process in \HI slab 1. 
  As the 
   \HII slab does not contribute to absorption and emission, $(I_{\nu}/\nu^{3})$ is invariant. 
Thus,  
\begin{align}
 I'_{\nu,1} & = b \nu^2\;\!  \left(\frac{1+z_2}{1+z_1}  \right)^3   
  \Big[ \;\! T_0  \;\! 
   \exp(-n_{\rm a}(z_1)\;\!\sigma_{\rm a}\;\!\phi(\nu-\nu^*_{\rm 21cm})
   \;\!\delta s) \nonumber \\
   & \hspace*{1.25cm}  
    +  T_1\;\! 
    \left[1- \exp(-n_{\rm a}(z_1)\;\!\sigma_{\rm a}\;\!\phi(\nu-\nu^*_{\rm 21cm})
    \;\!\delta s)\right] 
   \;\! \Big]  \ , 
\end{align} 
 where $\nu_{\rm 21cm}^* = \nu_{\rm 21cm}\;\! (1+z_2)/(1+z_1)$, and  
\begin{align} 
 I_{\nu,2}  
 & = b \nu^2\;\!
    \Big\{  \left(\frac{1+z_2}{1+z_1}  \right)^3     \exp(-n_{\rm a,2}\;\!\sigma_{\rm a}\;\!\phi(\nu-\nu_{\rm 21cm})
    \;\!\delta s ) 
     \nonumber \\ 
  & \hspace*{1.0cm} \times   \Big[\;\!  T_0  \;\! 
   \exp(-n_{\rm a,1}\;\!\sigma_{\rm a}\;\!\phi(\nu-\nu^*_{\rm 21cm})
   \;\!\delta s) \nonumber \\
   & \hspace*{1.5cm}  
    +  T_1\;\! 
    \left[\;\! 1- \exp(-n_{\rm a,1}\;\!\sigma_{\rm a}\;\!\phi(\nu-\nu^*_{\rm 21cm})
    \;\!\delta s)\right] 
    \Big]  \nonumber \\  
    & \hspace*{1cm} + T_2\;\! 
    \left[\;\! 1- \exp(-n_{\rm a,2}\;\!\sigma_{\rm a}\;\!\phi(\nu-\nu_{\rm 21cm})
    \;\!\delta s)\right] 
    \Big\} \ ,    
\end{align} 
  where $n_{\rm a,2}$ is the number density of absorber in \HI slab 2 
  and $T_2$ is the thermal temperature associated 
   with the line-formation process in \HI slab 2. 
Generally, $T_0 \neq T_1 \neq T_2$ 
 and $n_{\rm a,1} \neq  n_{\rm a,2}$.  
The spectrum of the radiation that emerges has a continuum component 
 and two line components, 
 one associated the absorption in \HI slab 1 and another associated  
  with the absorption in \HI slab 2. 
The continuum has a specific intensity 
\begin{align} 
I_{\nu,2} = b\nu^2 \left(\frac{1+z_2}{1+z_1} \right)^3 T_0 \ ,   
\end{align}   
 which is independent of the thermal conditions in the two \HI slabs.   
This is consistent with our assumption 
 that the two \HI slabs are optically thin to the continuum 
 and the \HII slab has no contribution to the radiation, 
 and that $\nu^{-3}I_\nu$ is an invariant quantity.   
For a sufficiently large difference between $z_1$ and $z_2$ 
  such that the two line-profile functions 
  do not overlap,  
the line associated with the first \HI slab is centred at 
  $\nu^*_{\rm 21cm} = \nu_{\rm 21cm}(1+z_2)/(1+z_1)$,  
  and the specific intensity at the line centre is  
\begin{align} 
I_{\nu,1} = b\nu^2 \left(\frac{1+z_2}{1+z_1} \right)^3 
  \big[\;\!(T_0 - T_1) \exp(-n_{\rm a,1}\;\!\sigma_{\rm a}\;\!\phi(0)
  \;\!\delta s) +T_1\;\! \big]  .    
\end{align}   
The line associated with the second \HI slab is centred at $\nu_{\rm 21cm}$, %ignoring terms with T1
  and the specific intensity at the line centre is 
\begin{align} 
I_{\nu,2} = b\nu^2 \Bigg\{ \bigg[ \left(\frac{1+z_2}{1+z_1} \right)^3 T_0 
   - T_2\;\!\bigg] \exp(-n_{\rm a,2}\;\!\sigma_{\rm a}\;\!\phi(0)
   \;\!\delta s) +T_2 \;\! \Bigg\} \ .    
\end{align}   
Here, it shows the two line components are not identical 
  with the correction of the shift in the line centre energies,  
  even for identical thermal conditions in the two \HI slabs 
  at time $t_0$, when the radiation enters the first \HI slab. 
Even when the micro-astrophysical processes, 
  such as heating and photon-pumping are absent,    
  cosmological expansion would 
  alter the temperature and the number density of the absorbers 
  in the slabs. 
If cosmological expansion is insignificant 
  over the interval when the radiation completes its journey 
  through the two \HI slabs and the \HII slab, 
 the spectrum has a continuum and only one line.  
The specific intensity of the continuum is simply   
\begin{align}   
 I_{\nu,2} & = b \nu^2 T_0 \ .  
\end{align}
The line is centred at frequency $\nu_{\rm 21cm}$,   
  and the specific intensity at the line centre is 
\begin{align}   
 I_{\nu,2} & = b \nu^2\;\! 
  \Big[\;\! (T_0-T_1)\;\!\exp(-(n_{\rm a,1} 
   +n_{\rm a,2})\;\!\sigma_{\rm a}\;\! \phi(0)
   \;\!\delta s)
    \nonumber \\ 
    &  \hspace*{1.5cm}
    +(T_1-T_2)\;\!\exp(-n_{\rm a,2}\;\!\sigma_{\rm a}\;\! \phi(0)
    \;\!\delta s)  + T_2\;\! \Big] 
  \ , 
\end{align} 
The line-centre specific intensity 
  becomes the same as the specific intensity 
  of the continuum neighbouring to the line,   
  i.e. the line vanishes, 
  when $T_2 = T_1 =T_0$.  
An emission line will result for $T_0 > T_1 > T_2$ 
    and an absorption line will result for $T_2 > T_1 > T_0$.  
These results are as expected 
  from the line-formation criteria.

The effect of cosmological expansion in the transport of ionising photons and the development of ionised cavities have been studied analytically and implemented in reionisation simulations \citep[see e.g.][]{Shapiro1987ApJ,Bisbas2015MNRAS,Weber2013AA,Fedchenko2018JPhCS}. The subsequent effects on 21-cm signals are also  discussed~\citep{Yu2005ApJ}. This cosmological effect is sometimes named as `light cone effect' and may lead to anisoptropy in the 21-cm power spectrum~\citep{Barkana2006MNRAS, Datta2012MNRAS, LaPlante2014ApJ}. However, this effect is generally ignored in  recent studies of 21-cm signals which are based on reionisation simulations.

% =====================
\section{Computational set-up for 21-cm radiative transfer calculation} 
\label{sec:setup}
% --------
We briefly recapitulate the C21LRT equation and the input default reionisation history model used throughout this paper first. The computational set-up for the radiative transfer calculations as the same as in~\citet{Chan2023MNRAS}, except for the inserted \HII zones.

The C21LRT equation in covariant form with our adopted FLRW cosmological model, when there is negligible scattering, is as follows,
\begin{align}
\label{eq:covariantlineRTstiemiInc} 
\frac{\rm d}{{\rm d} z} \left(\frac{I_{{\rm L},\nu}}{\nu^3}\right)
  & = (1+z)\;\!
    \Bigg[
    -\left(\kappa_{{\rm C},\nu}+ \kappa_{{\rm L}, \nu}\, \phi_{\nu}  \,[1-\Xi]\;\!
    \right)  
      \;\!\left(\frac{I_{{\rm L}, \nu}}{\nu^3}\right) 
       \nonumber \\  
     & \hspace*{2cm} 
       +  \frac{(\epsilon_{{\rm C}, \nu} 
     +\epsilon_{{\rm L}, \nu}\,\phi_{\nu})  }{\nu^3}\;\! 
     \Bigg] 
     \frac{\d s}{\d z}  \ , 
\end{align}
where the subscript ``${\rm L}$" and ``${\rm C}$" denote the 21-cm line and its neighbouring continuum (i.e. CMB in our calculations). 
Following their definitions in the previous section,  
$I_{{\rm L},\nu}$, $\kappa_{{\rm L},\nu}$ and $\epsilon_{{\rm L}, \nu}$ are the specific intensity, line absorption coefficient and line emission coefficient at 21-cm line centre, respectively. 
$\Xi$ is the factor for the stimulated emission. $\kappa_{{\rm C},\nu}$ and $\epsilon_{{\rm C}, \nu}$ are the continuum absorption coefficient and continuum emission coefficient and $\phi_{\nu}$ is the normalised line-profile functions. $s$ represents the photon's path length, and ${\d s}/{\d z}={c\d t}/{\d z}$ can be calculated based on our assumed cosmology (see Eqn.~(\ref{eq:dtdz})).

The line coefficients are calculated in a local rest frame (the same practice as in~\cite{Fuerst2004AA,Younsi2012AA,Chan2019MNRAS}) as follows,
\begin{align}  
  \epsilon_{\rm{L},\nu}  %& \equiv  \epsilon_{\rm{21cm},\nu}      \nonumber \\ 
    & = \frac{h\nu_{\rm{ul}}}{4\pi} \; 
      n_{\rm{u}} A_{\rm{ul}},  %\phi_{\nu, {\rm emi}} \ ,  
      \label{eq:21cmEmiCoefRT}  \\ 
\kappa_{{\rm L}, \nu} 
  &  = \frac{h \nu_{\rm ul}}{4\pi}
  \;\! n_{\rm l}B_{\rm lu}  =\frac{1}{8\pi}
  \left(\frac{c}{\nu_{\rm ul}} \right)^2 
  \left(\frac{g_{\rm u}}{g_{\rm l}}\right) 
\;\! n_{\rm l}A_{\rm ul},   \    
  \label{eq:normalisedLineAbs}  
\end{align}  
    where the subscript ``${\rm u}$" denotes the upper energy state 
     and ``${\rm l}$" denotes the lower energy state 
     of the \HI hyperfine transition. 
Here, $\nu_{\rm ul}=\nu_{\rm 21cm}, $ $g_{\rm u}$ and $g_{\rm l}$ 
  are the multiplicities (degeneracies) 
  of the upper and lower energy states, respectively, 
 $n_{\rm u}$ and $n_{\rm l}$ 
  are the number density of particles in the upper and lower energy states, respectively, 
 and $A_{\rm ul}$ and $B_{\rm ul}$ are the Einstein coefficients. We assume that the normalised line profile functions $\phi_{\nu,{\rm x}} \equiv \phi_{\rm x}(\nu -\nu_{{\rm line,0}})$,  with ${\rm x} \in\{{\rm abs},\; {\rm emi},\; {\rm sti}\}$  corresponding to absorption, spontaneous emission and stimulated emission, respectively. Then the factor for the stimulated emission is 
$\Xi  = ({n_{\rm{u}}g_{\rm{l}}})/
({n_{\rm{l}}g_{\rm{u}}})$.

With this C21LRT equation, the input \HI gas properties in EoR are the globally averaged spin temperature $T_{\rm s}(z)$ and ionised fraction $x_{\rm i }(z)$ from the `EOS 2021 all galaxies simulation result' (which used the 21CMFAST code)~\citep{Munoz2022MNRAS}. The density of \HI gas is calculated based on $x_{\rm i}$ and the adopted cosmological model. Then $n_{\rm u}$ and $n_{\rm l}$ are calculated with $T_{\rm s}$. We only consider CMB and leave other types of continuum emission to future studies. One 21-cm line profile is adopted for all redshifts throughout one calculation and it is parameterised as follows,
\begin{align}
\phi(\nu - \nu_{\rm 21cm}) 
  &  =\frac{1}{\sqrt{\pi} \;\! \Delta \nu_{\rm D}} \ 
{\rm exp} \left[-\left(
\frac{\nu-\nu_{\rm 21cm}}{\Delta \nu_{\rm D}}\right)^{2}\right] \  .  
\label{eq:Gaussian}
\end{align}
where 
\begin{align}
\Delta \nu_{\rm D}
  = \nu_{\rm 21cm}  
   \sqrt{\frac{2 k_{\rm B} T_{\rm k}}{m c^2} 
     + \left(\frac{v_{\rm turb}}{c}\right)^2} \       
\label{eq:effdopplerwidth}
\end{align} 
is the Doppler parameter.

As little do we know  
  about the turbulent velocity of HI gas, 
  from observations  
  or from reliable theoretical modelling,  
  in the ionisation front created by luminous sources,  
  we choose the turbulent velocity of gas in intra-cluster medium (ICM), IGM and inter stellar medium (ISM) as benchmarks. 
The characteristic $v_{\rm turb}$ for ICM is $\sim 100~{\rm km\,s^{-1}}$, created by large scale shocks
\citep{Subramanian2006MNRA,Hitomi2018PASJ,Basu2021Galax,Ruszkowski2011MNRAS,Schuecker2004AA,Parrish2010ApJ,Vazza2017MNRAS}. For IGM, turbulence can be caused by structure formation, galaxy merger and galactic outflows~\citep{Xu2020ApJ} with with $v_{\rm turb}\sim 1-10~{\rm km\,s^{-1}}$ reported in simulation studies \citep{Schmidt2015LRCA,Evoli2011MNRAS,Zhang2022FrASS} and also for observations of Lyman $\alpha$ forest~\citep{Bolton2022MNRAS}. 
For ISM, $v_{\rm turb}\sim 1-10~{\rm km\,s^{-1}}$ ~\citep{Oliva2018MNRAS,Patricio2018MNRAS} and can the turbulence can be driven by cold gas infall, gravitational instability or star formation activities~\citep{Patricio2018MNRAS}. We therefore adopt $v_{\rm turb}\sim 1,\, 10,\, 100~{\rm km\,s^{-1}}$ in our calculations. As will be demonstrated, the exact value of $v_{\rm turb}$ does not effect our conclusions. We then assume that $T_{\rm k}$ is always 0 in this paper~\footnote{To match $v_{\rm turb}=10~{\rm km s^{-1}}$, $T_{\rm k}=6060.67$ K which is already higher than the expected globally averaged temperature of \HI (\cf Figure 1 of ~\citet{Chan2023MNRAS}). Also, such a high temperature is usually accompanied with high ionisation fraction $x_{\rm i}$, diminishing the 21 cm signal from HI gas.}.
%----------------- after revision ---------------

%%%%%%%%%%%%%%%%%%%%%%%%%%%%%%%%%%%
%%%%%%%%%%%%%%%%%%%%%%%%%%%%%%%%%%%
\begin{figure}%fig4
    \centering
\includegraphics[width=1.0\columnwidth]{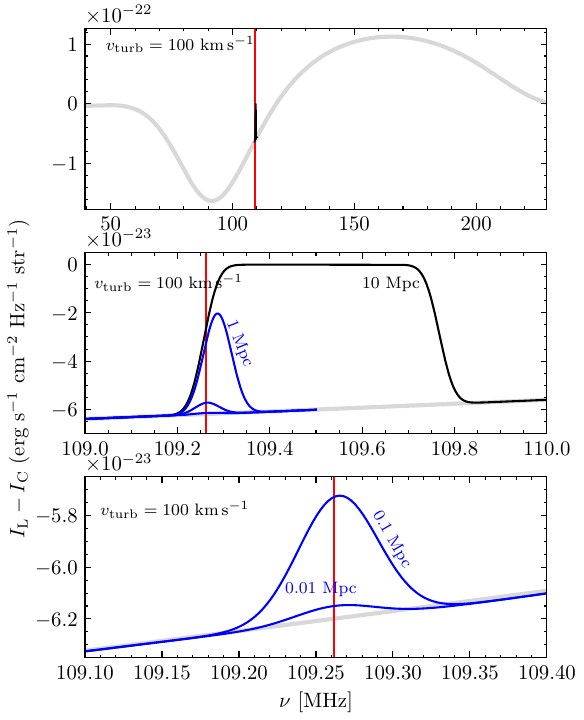}
    \caption{Features in
the 21-cm spectra at $z=0$ created by bubbles at $z_{\rm begin}=12$. All the spectra in this figure are calculated with  $v_{\rm turb}=100~{\rm km\,s^{-1}}$. The boundary of all the bubbles at the high redshift end $z_{\rm begin}=12.0$, which corresponds to $\nu=109.2620$ MHz, is marked with red vertical line in all three panels. 
%%%%
In the top panel, the globally averaged 21-cm spectrum without any bubble is plotted in thick grey line. The bubble features created by bubbles with diameters $d=0.01,\ 0.1,\ 1.0,\ 10.0$ Mpc are also plotted with think black lines in contrast. The bubble features are not distinguishable from each other in this panel and are shown in more details in the middle and bottom panels. $z=12.0$ is inside the absorption regime and all the bubble features can be understood as the decrease of the absorption signal. 
%%%%
In the middle panel, the features created by bubbles with $d= 10.0,\ 1.0,\ 0.1$ Mpc are plotted in black with solid, dashed and dash-dot lines. The original 21-cm spectra (in grey) and the features created by smaller bubbles (in blue) are also preserved in this panel for comparison. The transition between a bump and a valley happened when the bubble is $\gtrsim 10$ Mpc, this transition bubble size is mostly determined by $v_{\rm turb}$ as explained in the main text.
%%%%
In the bottom panel, the features created by bubbles with $d=0.01,\ 0.1$ Mpc are plotted in blue. The original spectra is plotted in grey but not visible as the curve representing $d=0.01$ Mpc is almost identical in this panel.}
\label{fig:bubble_z12_delta_v_1e7_cm_s}
\end{figure}
%%%%%%%%%%%%%%%%%%%%%%%%%%%%%%%%%%%
%%%%%%%%%%%%%%%%%%%%%%%%%%%%%%%%%%%

%%%%%%%%%%%%%%%%%%%%%%%%%%%%%%
\begin{figure}%fig5
\centering
\includegraphics[width=0.95\columnwidth]{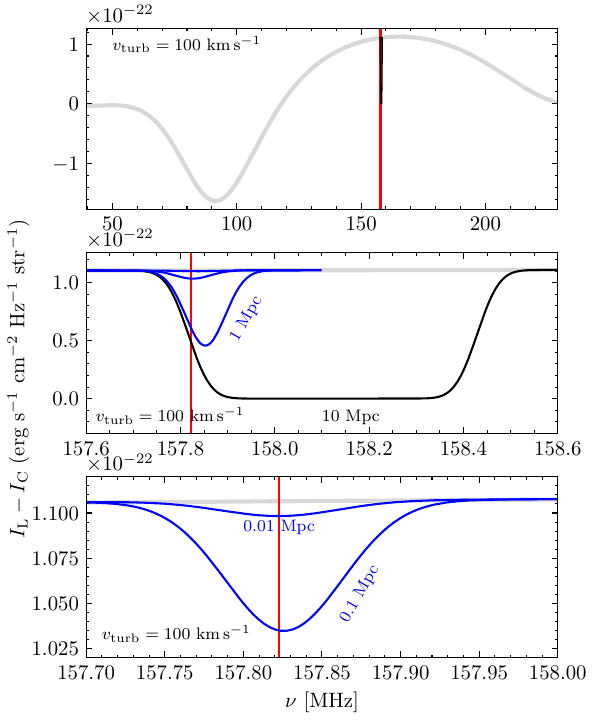}
\caption{Features in
the 21-cm spectra at $z=0$ created by bubbles at $z_{\rm begin}=8.0$. All the spectra in this figure are calculated with $v_{\rm turb}=100~{\rm km\,s^{-1}}$. The boundary of all the bubbles at the high redshift end $z_{\rm begin}=8.0$, which corresponds to $\nu=157.8229$ MHz, is marked with red vertical line in all three panels. $z_{\rm begin}=8.0$ is inside the emission regime and all the bubble features can be understood as decreasing of the emission signal. 
%%%%%%%%%%%%
In the top panel, the globally averaged 21-cm spectrum without any bubble is plotted in thick grey line. The bubble features created by bubbles with diameters $d=0.01,\ 0.1,\ 1.0,\ 10.0$ Mpc are also plotted with thin black lines in contrast. The bubbles features by larger bubbles and smaller bubbles are shown in more details in the middle and bottom panels. 
%%%%
In the middle panel, the features created by bubbles with $d= 10.0,\ 1.0,\ 0.1$ Mpc are plotted in black with solid, dashed and dash-dot lines. The original 21-cm spectra (in grey) and the features created by smaller bubbles (in blue) are also preserved in this panel for comparison. The transition between a dip and a valley happened when the bubble is $\gtrsim 10$ Mpc, this transition bubble size is mostly determined by $v_{\rm turb}$ as explained in the main text.
%%%%
In the bottom panel, the features created by bubbles with $d=0.01,\ 0.1$ Mpc are plotted in blue. The original spectra is plotted in grey but not visible as the curve representing $d=0.01$ Mpc is almost identical in this panel.}
\label{fig:bubble_z8_delta_v_1e7_cm_s}
\end{figure}
%%%%%%%%%%%%%%%%%%%%%%%%%%%%%%
\begin{figure}%fig6
    \centering
\includegraphics[width=1.0\columnwidth]{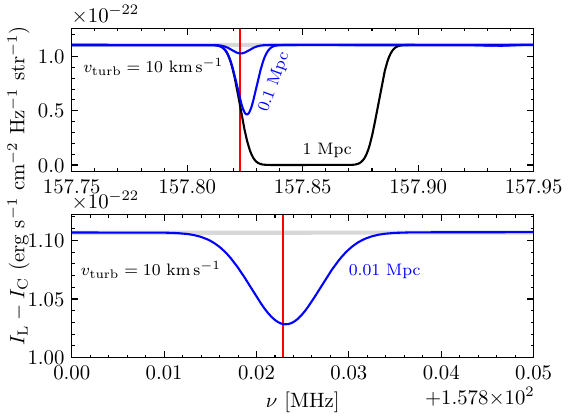}
    \caption{Features in
the 21-cm spectra at $z=0$ created by bubbles at $z_{\rm begin}=12$. All the spectra in this figure are calculated with $v_{\rm turb}=10~{\rm km\,s^{-1}}$. 
%%%%
In the top panel, the globally averaged 21-cm spectrum without any bubble is plotted in thick grey line (identical with some of the blues curved hence not visible). The bubble features created by bubbles with diameters $d=0.01,\ 0.1,\ 1.0$ Mpc are plotted in this panel. Due to a smaller $v_{\rm turb}$, the transition between a bump and a valley happened when the bubble is $\gtrsim 1$ Mpc.
%%%%
In the bottom panel, the features created by bubbles with $d=0.01,\ 0.1$ Mpc are plotted in blue. The original spectra is plotted in grey, although only visible in the bottom panel. The frequency $\nu=109.2620$ MHz corresponding to $z_{\rm begin}=12.0$ is marked with red vertical lines in both panels.}
\label{fig:bubble_z12_delta_v_1e6_cm_s}
\end{figure}

%%%%%%%%%%%%%%%%%%%%%%%%%%%%%%%%%%%%%%%%%%%%%%%%%%

More detailed description of the C21LRT formulation and the adopted default reionisation history can be found in ~\citet{Chan2023MNRAS}.

With the default reionisation history specified, we then insert \HII zones of spherical shapes (referred as `bubbles' in the following texts). Various methods have been employed in the literature to extract the topology and size distributions of ionised cavities from large scale simulation studies~\citep[e.g.][]{Furlanetto2006MNRAS,Shin2008ApJ,Friedrich2011MNRAS,Lin2016MNRAS}. These studies predict the distribution of sizes of ionised cavities and study the evolution of the distribution function throughout EoR. The characteristic sizes varies from $\lesssim 1$~Mpc at the beginning of EoR to $\gtrsim 10$~Mpc at the end of EoR~\citep{Iliev2006MNRAS,Shin2008ApJ,Friedrich2011MNRAS,Lin2016MNRAS}. 
To cover the possible sizes, we choose bubble diameters $d$ of $0.01,\ 0.1,\ 1.0,\ 10.0$~Mpc. Only one bubble is inserted in each scenario. For most of the scenarios, radiative transfer calculation is carried out along a single ray through the bubble centre. The comoving distance intercepted by the ray is $d$. We always fix the boundary of the bubble at higher redshift e.g. $z_{\rm begin}=12.0$ first. With the size ${\rm d}z$ of the bubble specified by $d$, we then calculate the boundary of the bubble at lower redshift $z_{\rm end}=z_{\rm begin}-{\rm d}z$. Along this ray, $n_{\rm HI}=0$ in the redshift cells between $z_{\rm begin}$ and $z_{\rm end}$. We insert bubbles in this way to facilitate the comparison of bubble features. We do not modify $T_{\rm s}$ in these redshift cells (as this should cause no difference). Each bubble in our calculation is resolved with at least 10 redshift cells.

\begin{figure}%fig7
    \centering
\includegraphics[width=1.0\columnwidth]{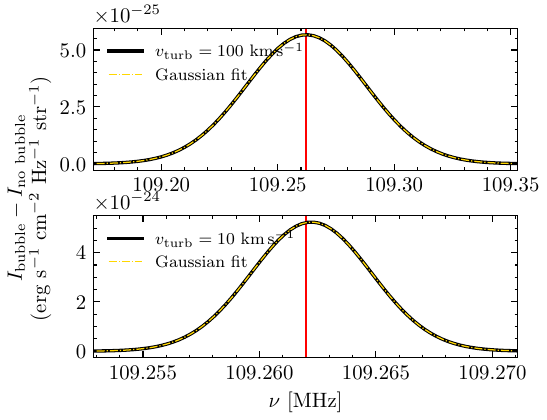}
\caption{Gaussian fit analysis of the features created by a $d=0.01$ Mpc bubble at $z=12.0$. In all three panels, the difference between the original spectra ($I_{\rm no bubble}$) and the spectra when there is a $d=0.01$ Mpc bubble at $z=12.0$ ($I_{\rm bubble}$) is plotted with black solid lines. The difference in spectra ($I_{\rm bubble}-I_{\rm no bubble}$-$\nu$) is then fitted with a Gaussian function, shown in yellow dashed curves. From top to bottom panel, the bubble features were calculated with $v_{\rm turb}=100,\ 10~{\rm km\,s^{-1}}$, respectively. As explained in the main text, the widths of Gaussian features are dominated by the line profiles because a $0.01$ Mpc bubble at $z=12.0$ is narrower compared to the widths of the line profiles, when measured in frequency. The amplitudes of the Gaussian features can be approximated as $I_{\rm bubble}-I_{\rm no bubble}\sim d\cdot \epsilon_{\rm L, \nu_{21cm}}$, see main text for details.}
\label{fig:fit_small_bubble_to_gaussian}
\end{figure}
%%%%%%%%%%%%%%%%%%%%%%%%%%%%%%%%%%%%%%%%%%%%%%%%%%  
%%%%%%%%%%%%%%%%%%%%%%%%%%%%%%%%%%%%%%%%%%%%%%%%%%  
Unless otherwise stated, 
 the maximum likelihood cosmological parameters obtained 
 by the \citet{PlanckXIII2016cosmopara} 
 are used in this work. 
At the present time, 
  the Hubble parameter is 
    $H_{0} = 100\,h_{0} = 67.74 ~\rm{km\;s^{-1}Mpc^{-1}}$, 
  the matter density is $\Omega_{\rm m,0} = 0.3089$, 
  the cosmological constant or vacuum density is 
     $\Omega_{\Lambda, 0}= 0.6911$, 
   and the radiation density is 
   $\Omega_{\rm r,0}=  4.1650\times 10^{-5}(h_0)^{-2}$ 
   \citep{Wright2006PASP}.

% =====================   
\section{Results and Discussions}   
\label{sec:resultsDiscussion}

% =====================
\subsection{Results}
\subsubsection{What determines the properties of bubble features}
We can first estimate the frequency range where the 21-cm spectra is affected by a bubble specified with $z_{\rm begin}$ and $z_{\rm end}$. Its effective size when measured in frequency (at $z=0$) ${\rm d}\nu_{\rm bubble}$ is
\begin{align}
{\rm d}\nu_{\rm bubble}=\nu_{\rm 21cm}[(1+z_{\rm end})^{-1}-(1+z_{\rm begin})^{-1}].
\end{align}
It is natural to assume that the spectrum will be affected in frequencies from $\nu_{\rm 21cm}(1+z_{\rm end})^{-1}$ to $\nu_{\rm 21cm}(1+z_{\rm begin})^{-1}$. We also expect the `neighbouring' frequencies to be affected due to the finite width of 21-cm line. The full width half maximum of the 21-cm line profile when measured with frequency in the local rest frame is 
\begin{align}
{\rm FWHM}_{\nu}& =2\;\! \sqrt{\ln{2}} \;\!
\left( \frac{v_{\rm turb}}{c} \right) \;\!
\nu_{\rm 21cm}  
\nonumber \\ 
& \approx{0.7889} \left( \frac{v_{\rm turb}}{100 ~{\rm km\,s^{-1}}}\right)~{\rm MHz}.
\label{eq:FWHM_nu}
\end{align}
Due to cosmological expansion, a 21-cm line with FWHM$_{\nu}$ at $z$ will become narrower, i.e.  $\nu_{\phi}={\rm FWHM}_{\nu}\frac{1}{1+z}$
when propagated to $z=0$. We therefore expect the spectrum to be affected at frequencies $\nu_{\rm 21cm}(1+z_{\rm end})^{-1}\pm{\rm d}\nu_{\phi}$ and $\nu_{\rm 21cm}(1+z_{\rm begin})^{-1}\pm{\rm d}\nu_{\phi}$. Note that we defined both ${\rm d}\nu_{\rm bubble}$ and ${\rm d}\nu_{\phi}$ at $z=0$ to facilitate the analysis of bubble features in the observed spectra, which are at $z=0$. 

This estimation is consistent with the C21LRT results. We first analysed scenarios at $z=12.0$ calculated with $v_{\rm turb}=100~{\rm km\,s^{-1}}$ as shown in \Fig\ref{fig:bubble_z12_delta_v_1e7_cm_s}. The original spectra without any bubble is plotted with thick grey line and the spectra with various bubble sizes in black in the top panel. In the middle panel, we zoomed into the relevant frequency range for the inserted bubbles. The smaller bubbles with $d=0.01,\ 0.1,\ 1$ Mpc are plotted in blue. Their effective widths are ${\rm d}\nu_{\rm bubble}=4.9572\times10^{-4},\ 4.9574\times10^{-3},\ 4.9670\times10^{-2}$~MHz, respectively, at $z=0$. These bubbles are narrower than the width of 21-cm line profile (${\rm d}\nu_{\rm bubble}\ll  {\rm d}\nu_{\phi}={0.06069}$ MHz), and the spectral width of the bubble feature (at $z=0$) ${\rm d}\nu_{\rm feature}$ is dominated by ${\rm d}\nu_{\phi}$. The effective size of the 10 Mpc bubble ($5.0018\times10^{-1}$~MHz) is larger than ${\rm d}\nu_{\phi}$. 
The corresponding bubble feature is plotted in black, with width dominated by ${\rm d}\nu_{\rm bubble}$. It is significantly broadened to be plateaus instead of bumps. The absolute value of specific intensity ($|I_{\rm L}-I_{\rm C}|$) is reduced to 0 at the plateau. In the bottom panel, we zoom in further to show the shape of the features created by smaller bubbles. Comparing the three curves calculated with $d=0.01,\ 0.1$ Mpc, we can see that whilst the widths of the bubble features are similar (dominated by ${\rm d}\nu_{\phi}$), the heights of these features increase as the sizes of the bubble increase.

The analysis of bubbles in the emission regime of the redshifted 21-cm global signal is similar. These bubbles have $d=0.01-10$ Mpc at $z_{\rm begin}=8.0$. Their features are also calculated with  $v_{\rm turb}=100~{\rm km\,s^{-1}}$ as shown in Fig.~\ref{fig:bubble_z8_delta_v_1e7_cm_s}. Bubbles manifest as dips or valleys (reduced emission) as shown in the top panel. We zoomed into the bubble region in the middle panel. The bubble features created by small bubbles ($d=0.01,\ 0.1,\ 1$ Mpc, ${\rm d}\nu_{\rm bubble}=5.9570\times10^{-4},\ 5.9572\times10^{-3},\ 5.9697\times10^{-2}$~MHz) have a dip like shape (in blue). 
The 10 Mpc bubble ($5.9989\times10^{-1}$ MHz) has effective size larger than ${\rm d}\nu_{\phi}=0.8766$~MHz and produce a valley-like shape. The transition between these two types of shapes is also determined by ${\rm d}\nu_{\phi}$. The features of smaller bubbles are enlarged in the bottom panel. Their depths also increase with the size of the bubble ${\rm d}\nu_{\rm bubble}$.

%%%%%%%%%%%%%%%%%%%%%%%%%%%%
%%%%%%%%gaussian table%%%%%%%%%%%
%%%%%%%%%%%%%%%%%%%%%%%%%%%%
%%%%%%%%%%%%%%%%%%%%%%%%%%%%
\begin{table*}
\centering
\caption{Results of Gaussian fit analysis. We fitted the bubble feature with a Gaussian function (see main text for the details of bubble feature and the fitting processes). From left to right, the columns are, 1 turbulent velocity $v_{\rm turb}$ adopted when calculating the bubble feature for $d=0.01$ Mpc at $z_{\rm begin}=12.0$, 2. fitted width at half maximum ${\rm d}\nu_{\rm fit}$ of the bubble feature, 3. adopted line profile width at half maximum
${\rm d}\nu_{\phi}$ as determined by $v_{\rm turb}$ at $z_{\rm begin}=12.0$, propagated to $z=0$, 4. difference between ${\rm d}\nu_{\phi}$ and ${\rm d}\nu_{\rm fit}$, 5. the maximum specific intensity (difference in intensity between the result with and without bubble) $I_{\max}$, 6. the fitted height of Gaussian $A_{\rm fit}$, 7. approximated height of the bubble feature calculated using 
Eqn.~\ref{eq:approx_bubble_feature}, 
8. relative difference between fitted Gaussian function and the calculated bubble feature averaged over from 
${(\nu_{\rm 21cm}-1.5 {\rm FWHM}_{\nu})}/{(1+z)}$ to 
${(\nu_{\rm 21cm}+1.5 {\rm FWHM}_{\nu})}/{(1+z})$.}
\label{tab:tab_gauss}
\begin{tabular}{cccccccc} % four columns, alignment for each
\hline
$v_{\rm turb}$ & ${\rm d}\nu_{\rm fit}$   & ${\rm d}\nu_{\phi}$ & ${\rm d}\nu_{\phi}-{\rm d}\nu_{\rm fit}$ & $I_{\max}$ &$A_{\rm fit}$ &$I_{\rm analytical}$  & $\overline{\delta}$ \\
${\rm km\;s^{-1}}$ & MHz&MHz&MHz& ${\rm erg\;(s\;cm^2\;Hz\;str)^{-1}}$&${\rm erg\;(s\;cm^2\;Hz\;str)^{-1}}$&${\rm erg\;(s\;cm^2\;Hz\;str)^{-1}}$&-\\
\hline
100&  $6.0696\times 10^{-2}$&  $6.0686\times 10^{-2}$&  $1.6087\times 10^{-4}$&$5.6694\times 10^{-25}$ &$5.6681\times 10^{-25}$&$7.5315\times 10^{-25}$&  $4.4682\times 10^{-5}$\\
10&  $6.0814\times 10^{-3}$&  $6.0686\times 10^{-3}$&  $2.0992\times 10^{-3}$& $5.2428\times 10^{-24}$ &$5.2396\times 10^{-24}$&$7.5315\times 10^{-24}$&  $3.4633\times 10^{-4}$\\
\hline
\end{tabular}
\end{table*}
%%%%%%%%%%%%%%%%%%%%%%%%%%%%
%%%%%%%%%%%%%%%%%%%%%%%%%%%%
%%%%%%%%%%%%%%%%%%%%%%%%%%%%

In the results above, we chose a large turbulent velocity ($v_{\rm turb}=100~{\rm km s^{-1}}$) to clearly demonstrate features created by the bubbles of $d=0.01-10$~Mpc, also the transitions from dip (peak) to valley (plateau) when bubble sizes increases in emission (absorption) regime. As the transition is determined by comparing ${\rm d}\nu_{\rm bubble}$ to ${\rm d}\nu_{\phi}$, the critical bubble size where the transition happens is determined by ${\rm d}\nu_{\phi}$ (which is determined by $v_{\rm turb}$ in our calculation). To demonstrate this dependence, we also calculated the bubble features at $z=12$ with $v_{\rm turb}=10~{\rm km\,s^{-1}}$, as shown in 
Fig.~\ref{fig:bubble_z12_delta_v_1e6_cm_s}. The line profile width for this $v_{\rm turb}$ is ${\rm d}\nu_{\phi}= 0.006069$~MHz. In the top panel, the transition now happens between $d=0.1$ Mpc and $d=1$ Mpc (the transition happens between $d=1$ Mpc and $d=10$ Mpc for $v_{\rm turb}=100~{\rm km\,s^{-1}}$ in the middle panel of Fig.~\ref{fig:bubble_z12_delta_v_1e7_cm_s}).

\subsubsection{Analytical approximations for bubble features}
For a $d=0.01$ Mpc bubble at $z=12.0$, its effective size ${\rm d}\nu_{\rm bubble}=4.9572\times10^{-4}$~MHz is smaller than the line profile width ${\rm d}\nu_{\phi}$ when adopting turbulent velocities $v_{\rm turb}=100,\ 10~{\rm km\,s^{-1}}$. The width of the bubble features calculated with these $v_{\rm turb}$ are therefore all dominated by ${\rm d}\nu_{\phi}$. We then fit the features created with $d=0.01$ Mpc with the line profile function (i.e. Gaussian functions in our calculations) for a more quantitative analysis. 

The fitting results in 
Fig.~\ref{fig:fit_small_bubble_to_gaussian} are for $v_{\rm turb}=100$ and $10~{\rm km\,s^{-1}}$ in the upper and lower panel. The bubble features calculated with C21LRT (black solid lines) and the fitted Gauasian functions (yellow dashed lines) are consistent over the frequency ranges from 
${(\nu_{\rm 21cm}-1.5 {\rm FWHM}_{\nu})}/{(1+z)}$ 
to ${(\nu_{\rm 21cm}+1.5 {\rm FWHM}_{\nu})}/{(1+z})$,  
where the fitting was carried out. The frequency corresponding to the high redshift boundary $z_{\rm begin}=12.0$ of all these bubbles are marked with red vertical lines. 
The fitted width ${\rm d}\nu_{\rm fit}$ and height $A_{\rm fit}$ of the Gaussian functions are listed in Table~\ref{tab:tab_gauss}. The fitted width ${\rm d}\nu_{\rm fit}$ for the two bubble features are very close to ${\rm d}\nu_{\phi}$ as determined by $v_{\rm turb}$. Also, $A_{\rm fit}$ is very close to the maximum specific intensity of the bubble feature $I_{\max}$. We calculated the relative difference between the bubble feature and the fitted Gaussian function and averaged it over from 
${(\nu_{\rm 21cm}-1.5 {\rm FWHM}_{\nu})}/{(1+z)}$ to 
${(\nu_{\rm 21cm}+1.5 {\rm FWHM}_{\nu})}/{(1+z)}$~\footnote{The frequency cells are equally spaced in log space and no weighting was added when calculating the average.}. The averaged relative difference $\overline{\delta}$ for the two bubble features are all less than $10^{-4}$. The small values of ${\rm d}\nu_{\phi}-{\rm d}\nu_{\rm fit}$, $I_{\max}-A_{\rm fit}$ and $\overline{\delta}$ shows that the bubble features are very close to Gaussian functions and the bubble width is well approximated with ${\rm d}\nu_{\phi}$.

We then analyse the height of the feature $A_{\rm fit}$ of these bubble features. As discussed above, $A_{\rm fit}$ grows with ${\rm d}\nu_{\rm bubble}$. We also expect it to be proportional to the 21-cm emission coefficient~\footnote{The 21-cm absorption coefficient in C21LRT is determined by the correction due to stimulated emission, which is much smaller than the emission coefficient.}. We found that $A_{\rm fit}$ can be approximated as 
\begin{equation}
\label{eq:approx_bubble_feature}
A_{\rm fit}\approx C_{\text{convolution}}\cdot I_{\rm analytical}=C_{\text{convolution}}\cdot C_{{\rm d}\nu}\cdot C_{\rm redshift}\cdot  d\cdot \epsilon_{\rm L, \nu_{21cm}},
\end{equation}
where $C_{\rm redshift}=(1+z)^3$ results from cosmological expansion (we may view this the factor as the $\nu^{-3}$ factor in the expression of the invariant intensity). $C_{{\rm d}\nu}=1+z$ is caused by the width of the relevant integration frequency range scales as $1+z$ when the features propagate from $z$ to the observer on earth. The $C_{\text{convolution}}$ factor resulted from convolution effect of line broadening in the local rest frames and the propagation of radiation. The values of $I_{\rm analytical}$ for the three bubble features are also listed in Table~\ref{tab:tab_gauss}. By comparing $A_{\rm fit}$ with $I_{\rm analytical}$, the value of $C_{\text{convolution}}$ is approximately $0.75$.

These three examples are calculated with fixed ${\rm d}\nu_{\rm bubble}$ and different ${\rm d}\nu_{\phi}$. The results show that when the bubble feature is dominated by the line profile (${\rm d}\nu_{\rm bubble}\ll  {\rm d}\nu_{\phi}$), the width of its spectral signature is well approximated by ${\rm d}\nu_{\phi}$. The height of the bubble's spectral feature is proportional to ${\rm d}\nu_{\rm bubble}/{\rm d}\nu_{\phi}$ and can be approximated with Eqn.~\ref{eq:approx_bubble_feature} . This analysis also applies to other small bubbles. For example, at a given redshift the bubble feature produced with $d=0.1$ Mpc calculated with $v_{\rm turb} =100~{\rm km\;s^{-1}}$ is comparable to that produced with $d=0.01$ Mpc calculated with $v_{\rm turb} =10~{\rm km\;s^{-1}}$.

The large bubbles can be approximated in a similar way. The features created by large bubbles (which satisfy ${\rm d}\nu_{\rm bubble}\gg  {\rm d}\nu_{\phi}$) can be separated into three part. For example, the 10 Mpc bubble feature in the middle panel of Fig.~\ref{fig:bubble_z12_delta_v_1e7_cm_s} starts and ends approximately at frequencies
\begin{align}
\label{eq:nubegin}
\nu_{\rm begin}=(\nu_{\rm 21cm}-0.5{\rm FWHM}_{\nu})/(1+z)|_{z=z_{\rm begin}}
\end{align}
and 
\begin{align}
\label{eq:nuend}
\nu_{\rm end}=(\nu_{\rm 21cm}+0.5{\rm FWHM}_{\nu})/(1+z)|_{z=z_{\rm end}}.
\end{align}
This features can be approximately separated by 
\begin{align}
\label{eq:nuleft}
\nu_{\rm left}=(\nu_{\rm 21cm}+0.5{\rm FWHM}_{\nu})/(1+z)|_{z=z_{\rm begin}}
\end{align}
and 
\begin{align}
\label{eq:nuright}
\nu_{\rm right}=(\nu_{\rm 21cm}-0.5{\rm FWHM}_{\nu})/(1+z)|_{z=z_{\rm end}}
\end{align}
into left, middle and right part. The left part (from $\nu_{\rm begin}$ to $\nu_{\rm left}$) and right part (from $\nu_{\rm right}$ to $\nu_{\rm end}$) are both approximately half of a Gaussian shape and their width are approximately $0.5{\rm FWHM}_{\nu}/(1+z)=0.5{\rm d}\nu_{\phi}$, the intensity of the middle part (from $\nu_{\rm left}$ to $\nu_{\rm right}$) is very close to 0. These estimated frequencies are of $\lesssim 0.1$~MHz accuracy. For example, $\nu_{\rm left}=109.2923$~MHz for the 10 Mpc bubble feature in Fig.~\ref{fig:bubble_z12_delta_v_1e7_cm_s}; the bubble feature ($I_{\rm bubble}-I_{\rm no bubble}$) computed from C21LRT calculation has the (left) maximum value at $109.3358$~MHz.

For bubbles of intermediate size (${\rm d}\nu_{\rm bubble}\sim  {\rm d}\nu_{\phi}$), they produce spectral features between $\nu_{\rm left}$ and $\nu_{\rm right}$. Their spectral features are similar to a Gaussian function but can not be fitted perfectly with a Gaussian function. This is because their spectral features are determined by the convolution between the line profile and other radiative transfer effects over a relatively wide redshift range~\footnote{When we fit these bubble features with Gaussian function, the fitted Gaussian function is flatter than the bubble feature, i.e. the height of bubble feature $|I_{\max}|$ is noticeable larger than the fitted height $|A_{\rm fit}|$.}. 

We note that these approximations are only valid because we adopted a constant $v_{\rm turb}$, hence a constant normalised line profile, for each calculation and all the bubbles we inserted are fully ionised. The features will be too complicated to be approximated analytically if we adopt  realistic $n_{\rm HI}$ and $v_{\rm turb}$ values, which could have large spatial variation within one ionised cavity created by an astrophysical source.

%%%%%%%%%%%%%%%%%%%%%%%%%%%%%%%%%%%%%%%%%%%%%%%%%%

%%%%%%%%%%%%%%%%%%%%%%%%%%%%%%%%%%
%%%%%%%%%%%%%%%%%%%%%%%%%%%%%%%%%%
\begin{figure}
\centering
\includegraphics[width=0.8\columnwidth]{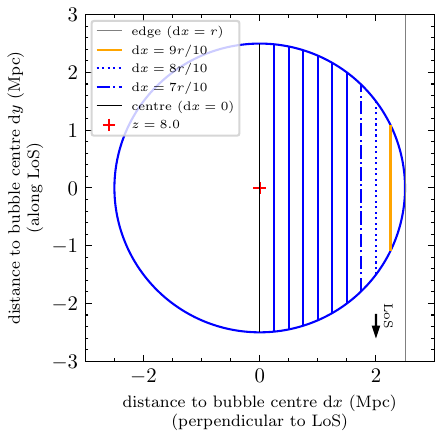}
\caption{Using ten rays to resolve a bubble with $r=2.5$~Mpc at $z=8$. The first ray is determined by the bubble centre and the observer on earth (d$x=0$), plotted in black. All the other 9 rays are parallel to the first ray, with distance d$x=r/10,\ 2r/10,\ ......, 9r/10$. We use d$x$ to denote distance to the first ray. The edge of the bubble is marked by the thick grey line (d$x=r$), where the ray does not intersect the \HII zone. }
\label{fig:resolve_bubble_z8_R25Mpc_input}
\end{figure}

\begin{figure}
\centering
\includegraphics[width=1.0\columnwidth]{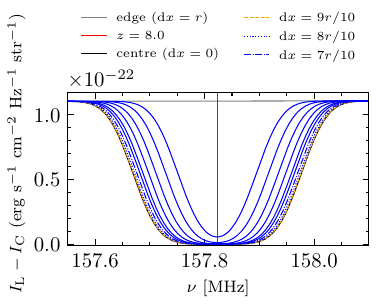}
\caption{21-cm spectra at $z=0$ calculated along the ten rays which intersect a bubble with $r=2.5$~Mpc at $z=8$. The ten rays that intersect the bubble are as specified in Fig~\ref{fig:resolve_bubble_z8_R25Mpc_input}.}
\label{fig:resolve_bubble_z8_R25Mpc}
\end{figure}
%%%%%%%%%%%%%%%%%%%%%%%%%%%%%%%%%%
%%%%%%%%%%%%%%%%%%%%%%%%%%%%%%%%%%

\subsubsection{Resolve a bubble with multiple rays}
\label{subsubsec:multiple_rays}
When ionised bubbles are large enough, it is possible to resolve each of them with multiple rays and study the changes in 21-cm features due to their spatial variations.

We used ten rays to resolve a bubble with $d=5.0$~Mpc ($r=2.5$~Mpc) at $z=8$. Different to the previous calculations, here we set the bubble centre at $z=8$ and calculated $z_{\rm begin}$ and $z_{\rm end}$ for each ray. The bubble and the ten rays are illustrated in 
\Fig\ref{fig:resolve_bubble_z8_R25Mpc_input}. The first ray is determined by the bubble centre and the observer on earth (d$x=0$), which also determines the LoS. All the other 9 rays are parallel to the first ray, with distance d$x=r/10,\ 2r/10,\ ......, 9r/10$. We use d$x$ to denote distance to bubble centre in the perpendicular direction with respect to LoS. The edge of the bubble is marked by the thick grey line (d$x=r$), where the ray does not intersect the \HII bubble. We adopted a turbulent velocity of 100 ${\rm km\;s^{-1}}$~\footnote{We note that this 100 ${\rm km\;s^{-1}}$ turbulent velocity may not be achieved for a realistic ionisation bubble. It is chosen such that the distortion of the apparent shape of the ionisation front, due to the finite speed of light, can be clearly demonstrated.}.

%%%%%%%%%%%%%%%%%%%%%%%%%%%%%%%%%%%%%%%%%%%%%%%%%%
\begin{figure}%fig10
\centering
\includegraphics[width=0.8\columnwidth]{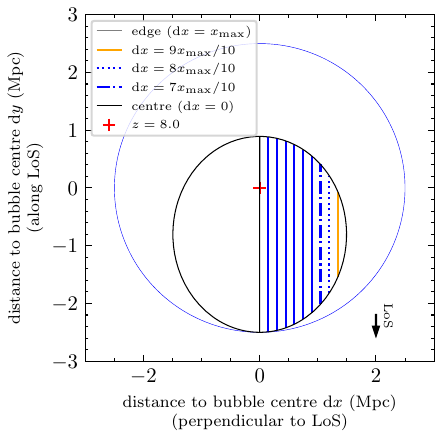}
\caption{Using ten rays to resolve an expanding bubble at $z=8$ whose ionisation front is expanding with a velocity of $v=0.9~c$ in the local rest frame of an assumed ionising source, marked by the red cross. When the bubble has expanded to $r=2.5$ Mpc in the rest frame (marked by the blue circle), we calculate the apparent shape due to finite speed of light (see main text for details). The apparent shape is indicated with the black ellipse. The ten rays are equally spaced, with the first ray intercept the bubble centre (d$x=0$), plotted in black.
The other 9 rays have distance d$x=x_{\max}/10,\ 2x_{\max}/10,\ ......, 9x_{\max}/10$. The edge of the bubble is marked by the thick grey line  (d$x=r$), where the ray does not intersect the \HII zone. }
\label{fig:resolve_expand_bubble_z8_R25Mpc_input}
\end{figure}
\begin{figure}
\centering
\includegraphics[width=1.0\columnwidth]{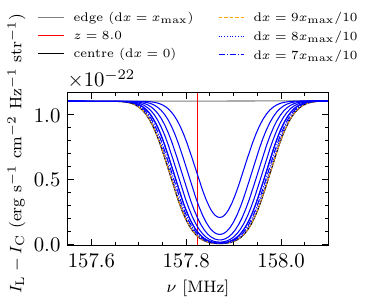}
\caption{21-cm spectra at $z=0$ calculated along the ten rays which intersect an expanding bubble with the ionisation front's velocity of $v=0.9~c$ and an initial radius $r=2.5$~Mpc in the local rest frame of the source at $z=8$. The ten rays that intersect the bubble are as specified in Fig~\ref{fig:resolve_expand_bubble_z8_R25Mpc_input}. 
Compared to the spectra shown in Fig~\ref{fig:resolve_bubble_z8_R25Mpc} for a non-locally expanding bubble, the spectral features are narrower and are no longer symmetrical with respect to the frequency marked by the red vertical line ($z=8.0)$.
}
\label{fig:resolve_expand_bubble_z8_R25Mpc}
\end{figure}
%%%%%%%%%%%%%%%%%%%%%%%%%%%%%%%%%%%%%%%%%%%%%%%%%%
%%%%%%%%%%%%%%%%%%%%%%%%%%%%%%%%%%%%%%%%%%%%%%%%%%  

The 21-cm spectra at $z=0$ for the bubble region are shown in \Fig\ref{fig:resolve_bubble_z8_R25Mpc}. The spectra are coded with the same color and line styles as the corresponding rays in \Fig\ref{fig:resolve_bubble_z8_R25Mpc_input}. The bubble features grows in both depth and width from d$x=9r/10$ to d$x=0$. Each bubble feature appears symmetrical to the frequency that corresponds to the bubble centre at $z=8.0$. When d$x=9r/10 $, the distance intercepted by the ray is small, and the bubble feature is still dominated by line profile width. As d$x$ decreases, the intercepted distance grows and dominates over the line profile width.  

So far, we have been measuring the bubble size or the intercepted distance in the rest frame of the (assumed) ionising source at the centre of the bubbles. This is no longer accurate if the bubble is expanding with velocities $v$ comparable to the speed of light. When $v$ is comparable to $c$, the size of the bubble changes significantly within the light crossing time (e.g. of the ray which intersect the bubble centre) as discussed in Sec.~\ref{subsec:cosmological expansion}. The apparent shape probed with 21-cm line is different from the shape observed from the local rest frame of the ionising source. We followed~\citet{Yu2005ApJ} and calculated the apparent shape of a bubble with a constant $v=0.9~c$, as shown with the black ellipse in \Fig\ref{fig:resolve_expand_bubble_z8_R25Mpc_input}. Instead of assigning rays by d$x=r/10 ,\ 2r/10 , ... r$, we first calculate the maximum distance to the centre ray $x_{\max}$ and assign rays with d$x=x_{\max} /10 ,\ 2 x_{\max}/10 , ... 9x_{\max}/10$. We can see that due to a large $v$, the apparent shape significantly changed~\footnote{The bubble is still axisymmetric with respect to the ray which intercept the bubble centre.}. The intercepted distances decreased for all the ten rays compared to \Fig\ref{fig:resolve_bubble_z8_R25Mpc_input}. The resulting 21-cm spectra $z=0$ are shown in \Fig\ref{fig:resolve_expand_bubble_z8_R25Mpc}. The rays close to the edge of bubbles intercept much shorter distances (where there is no \HI) and the width of the corresponding bubble features are dominated by line profile width. All the spectra features are narrower compared to those in \Fig\ref{fig:resolve_bubble_z8_R25Mpc}. These spectra no longer appear symmetrical with respect to the frequency marked with red vertical line ($z=8.0$).

\subsubsection{Differences between C21LRT and optical depth parametrisation}\label{subsubsec:diffC21LRT_opdepth}

Using identical HI gas properties as inputs, 
the absolute difference in the spectra at $z=0$ calculated with C21LRT and optical depth parmetrisation method, denoted by $\delta$, primarily arises from two factors. 
First, the optical depth method tracks the signal evolution only in redshift space, while C21LRT tracks 21-cm line in a two-dimensional (redshift and frequency) space.
This results in an approximate $5-10\;\!\!\%$   difference across all redshifts. 
The second factor stems from the simplifications used in deriving the optical depth parametrisation, at the expense of neglecting small-scale variations of HI gas. Both factors are discussed in \citet{Chan2023MNRAS}.
The second factor is evident when HI gas properties undergo abrupt changes, such as the transition between neutral and ionised zones. 
For a non-zero FWHM$_{\nu}$, $\delta$ reaches its maximum at approximately $\nu_{\rm left}$ and $\nu_{\rm right}$ as defined in Eqns.~\ref{eq:nuleft} and~\ref{eq:nuright}, 
where $I_{\rm L}-I_{\rm C}$ immediately drops to 0 based on optical depth method, whereas 
$|\delta|\approx (I_{\rm L}-I_{\rm C})/2$ 
based on C21LRT calculations\footnote{This means that, compared to the C21LRT results (considered more accurate), the signals computed from the optical depth method can have 100~\% discrepancy near ionisation fronts.}. 
For smaller bubbles where ${\rm d}\nu_{\phi}>{\rm d}\nu_{\rm bubble}$, 
$|\delta|$ remains greater than 0 throughout the relevant frequency range (from $\nu_{\rm begin}$ to $\nu_{\rm end}$, as defined in Eqns.~\ref{eq:nubegin} and 
\ref{eq:nuend}). 
For larger bubbles, $|\delta|$ decreases to 0 between $\nu_{\rm left}$ and $\nu_{\rm right}$. 
These differences mostly result from the rest frame 21-cm broadening, with a minor contribution from radiative transfer effects.

The bubble features can also be understood as follows: For the optical depth method (with $v_{\rm turb}=0$ implicitly assumed), the `loss' of the 21 cm signal due the presence of an ionised bubble is distributed between $\nu_{\rm 21cm}/(1+z)|_{z=z_{\rm begin}}$ and $\nu_{\rm 21cm}/(1+z)|_{z=z_{\rm end}}$. 
For the cases presented in Figs. 4, 5, 6, 7, 
the value of $v_{\rm turb}$ determines the ${\rm FWHM}_\nu$ (eq.~\ref{eq:FWHM_nu}). 
The `loss' of 21 cm signal is distributed over a broader frequency range between $\nu_{\rm begin}$ and $\nu_{\rm end}$. 
The shape of the spectral feature is determined mainly by the 21-cm line profile. 

Additionally, 
when using the optical depth method to compute the 21-cm signals seen by a present-day observer, it often involves a separate prescription to incorporate the light from past epochs that can reach us now. This is often done by constructing a light cone from sequential snapshots of the simulated box, where the same part of the universe has evolved with cosmological time. 
However, the optical depth method 
is inaccurate 
when 21-cm line is broadened or not optically thin \citep{Chapman2019MNRAS}. 
It can also overlook certain cosmological effects, such as the alterations in the apparent shape of the expanding ionisation fronts. 

% =====================
\subsection{Discussion} 
\subsubsection{Implications of our results}

For more sophisticated ionised cavities in cosmological simulations, their impacts on the 21-cm spectra and power spectra are currently all calculated with the optical depth parameterisation. This means that the radiative processes (e.g. line broadening due to turbulent velocity) which happens on length scales of or smaller than the sizes of the ionised cavities are missing in their calculation. When observational data become available, the sizes of ionised cavities in EoR inferred from their calculated 21-cm power spectra can have large uncertainties. For example, if ionised cavities all have sharp boundaries (the ionised fraction changes quickly from 1 to 0) and have a characteristic diameter $d$, the optical depth parametrisation would predict a sharp change in the power spectra near $k=1/d$, while full radiative transfer calculations would predict gradual changes near $k=1/d$.

\subsubsection{Some remarks on the 
ionised cavity models}

In this work 
we have used the two-zone model 
to show the features that it imprints on 
the 21-cm spectra. The two-zone model  
  is a plausible representation 
  of the ionised cavities carved 
  out by the very luminous astrophysical sources when the partially ionised zone has negligible size, 
  such as quasars~\footnote{We tested the three-zone model by adding a thin partially zone near the boundary (similar to the illustration in Figure 1)  and found that the features in the 21-cm spectra 
  of the two-zone and three-zone models 
  are actually very similar. 
  We therefore do not show the results 
  of three-zone models in this paper.}. The propagation of the ionising front created by powerful ionising sources, as captured in the expanding bubble scenario (presented in Sec~\ref{subsubsec:multiple_rays}) 
has demonstrated that 
the time evolution of ionisation structures needs to be corrected.

In some situations, 
  the two-zone model 
  may not be able to fully captures 
  thermodynamics and geometrical   
  properties of the ionised cavities. 
For instance,  
  UV radiation from 
  individual sources such as Population III stars or Population II stars 
  would ionise the surrounding \HI gas 
  gradually over time.  
This would allow 
  the photoionisation and recombination processes 
  to reach a quasi-equilibrium state,   
  in which 
  the size of the cavity  
  would be approximately the 
  Str{\"o}mgren radius~\citep[see][]{Stromgren1939ApJ}. 
In the more complicated situations, 
such as   
  that the ionisation source  
  is a group of star-forming galaxies 
  or a galaxy hosting AGNs,  
  the radiation flux and spectrum 
  would vary with time. 
The interplay between microscopic processes, 
  such as photoionisation, recombination 
  and radiative cooling loss, 
  and macroscopic processes, 
  such as hydrodynamical bulk flows, 
  will give rise to instabilities.   
The distribution of \HI gas near the ionising sources 
 is no longer uniform and smooth.  
The media both inside and outside the ionised cavities
  can be clumpy,
  and 
  the 21-cm spectral signature of these cavities  
  will be very different 
  to those obtained from the model bubbles 
  constructed using a globally averaged $n_{\rm HI}$. The development of the ionisation structure created by these sources are jointly determined by the ionisation, recombination processes as well as the thermodynamic processes~\citep{Axford1961RSPTA,Newman1968ApJa,Yorke1986ARAA,Franco2000ApSS}. 
Transition layer of partially ionised gas can vary in thickness, ionisation state and thermodynamics properties. 
Furthermore, in the presence of X-ray emissions, which  
penetrate more deeply than UV photons and are more efficient at heating gas, ionised cavities are expected to much larger with a hazier boundary. 
Over time, ionised bubbles also came to overlap. All these factors are expected to individually and collectively affect the 21-cm power spectra ~\citep{Finlator2012MNRAS,Kaurov2014ApJ,Hassan2016MNRAS,Mao2020MNRAS}. Our future studies will look into these complex ionisation structures in more detail, as C21LRT can interface with the simulated data with small scale ($\lesssim 10$~kpc) structures in the future
(see Appendix~\ref{sec:computational_time} 
for details).

%%%%%%%%%%%%%%%%%%%%%%%%%%%%%%%%%%%%%%%%%%%%%%%%%%
%%%%%%%%%%%%%%%%%%%%%%%%%%%%%%%%%%%%%%%%%%%%%%%%%%   
% =====================   
\section{Conclusion}\label{sec:conclusion}
We investigated the 21-cm spectral features imprinted by individual spherical ionised cavities enveloped by \HI medium, adopting a 
covariant formulation for tracking 21-cm signals in both redshift and frequency space. We studied the improvement in accuracies of the 21-cm signals, compared to adpoting the optical depth parametersation, which tracks 21-cm in the redshift space only.

We use a cosmological radiative transfer code C21LRT for the numerically calculation of 21-cm signals. We showed that the evolution of ionised cavities as driven by ionisation, thermodynamical and cosmological effects, would affect the apparent shape of the cavities 
when probed with 21-cm line. 
The apparent shape of an evolving cavity is different from its shape in the rest frame of its stationary cavity centre, 
as there is a time lap 
between the radiation from 
the farside and from the nearside of the cavity. 

We employed a set of single-ray calculations 
to show how the spectral features evolve with various bubble diameters $d=0.01,\ 0.1,\ 1.0,\ 10.0$ Mpc. 
We found that the widths of spectral features imprinted by bubbles are jointly determined by the 21-cm FWHM$_{\nu}$, the bubble diameter $d$ and the redshift of the bubble $z$. The spectral feature width at redshift $z$ can be approximated with max$({\rm d}{\nu_\phi},{\rm d}\nu_{\rm bubble})$, where ${\rm d}{\nu_\phi}=$FWHM$_{\nu}/(1+z)$ and ${\rm d}\nu_{\rm bubble}$ is the bubble diameter when measured in frequency space at $z=0$.

When the 21-cm line FWHM$_{\nu}$ dominates (${\rm d}\nu_{\rm bubble}\ll  {\rm d}\nu_{\phi}$), as we adopted a Gaussian function shape 21-cm line profile, the bubble spectral feature has a Gaussian function like shape. For bubbles with smallest $d$ in our paper (0.01 Mpc), their spectral features can be fitted with Gaussian function with negligible residuals. As $d$ increases (${\rm d}\nu_{\rm bubble}\sim  {\rm d}\nu_{\phi}$), the convolution between line profile and other radiative transfer and cosmological effects becomes non-negligible and their spectral shape deviates from Gaussian functions. For these bubbles (${\rm d}\nu_{\rm bubble}\lesssim {\rm d}\nu_{\phi}$), the 21-cm specific intensity $I_{\rm L}-I_{\rm C}$ does not decrease to 0 in the corresponding frequency ranges.

For large bubbles  (${\rm d}\nu_{\rm bubble}\gg  {\rm d}\nu_{\phi}$), the widths of their spectral features are dominated by ${\rm d}\nu_{\rm bubble}$. These features can be divided into three parts. The shape of the left and right parts of the spectral feature is similar to half Gaussian functions. They are produced by the transition between \HI and \HII zone and their shape is mainly determined by the 21-cm line profile. The middle part of their spectral feature has $I_{\rm L}-I_{\rm C}=0$. In comparison, the optical depth parametersation predicts that 21-cm signal will diminish to 0 for a fully ionised bubble, regardless of its size. We showed that the 21-cm intensities computed with it can have large discrepancies in the transition zones (ionisation fronts) from those computed with C21LRT.

The 21-cm signals associated with length scales equal to or smaller than the sizes of the ionized cavities need to be tracked in both redshift and frequency space for accuracy. For these scales, physical processes, such as line-broadening due to turbulent motion of the gas imprints on the 21-cm signals, can not be accurately computed with the  optical depth parametrisation.
Explicit covariant radiative transfer, such as the C21LRT, 
is necessary 
for correctly and self-consistently accounting for 
the convolution of local 
(thermodynamics and atomic processes 
and bubble dynamics) 
and global (cosmological expansion) effects onto the radiation that we receive from EoR.

%%%%%%%%%%%%%%%%%%%%%%%%%%%%%%%%%%%%%%%%%%%%%%%%%%
%%%%%%%%%%%%%%%%%%%%%%%%%%%%%%%%%%%%%%%%%%%%%%%%%%   
% =====================
\section*{Acknowledgements}
% =====================

We thank Richard Ellis for critically reading 
  through the manuscript. 
JYHC is supported by the University of Toronto Faculty of Arts \& Science Postdoctoral Fellowship with the Dunlap Institute, and 
 the Natural Sciences and Engineering Research Council 
 of Canada (NSERC), [funding reference \#CITA 490888-16], 
 through a CITA Fellowship. 
  The Dunlap Institute is funded through an endowment established by the David Dunlap family and the University of Toronto. 
QH is supported by a UCL Overseas Research Scholarship 
  and a UK STFC Research Studentship.    
KW and QH acknowledge 
  the support from the UCL Cosmoparticle Initiative. 
This work is supported in part 
  by a UK STFC Consolidated Grant awarded to UCL-MSSL. 
This research had made use of NASA’s Astrophysics Data System.

%%%%%%%%%%%%%%%%%%%%%%%%%%%%%%%%%%%%%%%%%%%%%%%%%%
\section*{Data Availability}
The theoretical data generated in the course of this study are available from the corresponding author QH, upon reasonable request.

%%%%%%%%%%%%%%%%%%%%%%%%%%%%%%%%%%%%%%%%%%%%%%%%%%



\bibliographystyle{mnras}
\bibliography{main} % if your bibtex file is called example.bib

%%%%%%%%%%%%%%%%%%%%%%%%%%%%%%%%%%%%%%%%%%%%%%%%%%

%%%%%%%%%%%%%%%%% APPENDICES %%%%%%%%%%%%%%%%%%%%%

%\newpage
\appendix

%%%%%%%%%%%%%%%%%%%%%%%%%%%%%%%%%%%%%%%%%%%%%%%%%%
%%%%%%%%%%%%%%%%%%%%%%%%%%%%%%%%%%%%%%%%%%%%%%%%%%

\section{computational time}
\label{sec:computational_time}
In our calculations, \HII bubbles are all resolved with at least 10 redshift cells. We still used a rectangular two dimensional grid in $\log{\nu}$--$\log(1+z)$ space (see Appendix A of~\cite{Chan2023MNRAS} for details).
For the smallest bubbles $d=0.01$ Mpc, their sizes when measured in redshift space are $\sim 10^{-5}$ at $z\sim 10$, hence we have already used redshift resolution of $\Delta_{\log{(1+z)}} \equiv \log{(1+z')} - \log{(1+z)}<10^{-6}$. When measured with comoving distances, this shows that C21LRT can resolve scales down to $\sim 1$ kpc. The best frequency resolution in this paper, when adopting the smallest turbulent velocity of $1~{\rm km\;s^{-1}}$ to produce the result shown in Fig.\ref{fig:fit_small_bubble_to_gaussian} , is $\Delta_{\log{\nu}} \equiv \log{\nu_{z'}}-\log{\nu_{z}}=10^{-6}$. This 
shows that C21LRT can resolve small turbulent velocity $v_{\rm turb}=10~{\rm km\;s^{-1}}$ with accuracy. In each calculation, the ray is traced from $z=z_{\rm begin}+\Delta z_1$ to $z=z_{\rm end}-\Delta z_2$, where $\Delta z_1$ and $\Delta z_2$ need to be larger than $\Delta z'_1$ and $\Delta z'_2$. Their values are determined by the effective line profile width (when translated into redshift space) as follows,
\begin{align}
\frac{\nu_{\rm 21cm}}{1+z_{\rm begin}+\Delta z'_1}=\frac{\nu_{\rm 21cm}-{\rm FWHM}_{\nu}}{1+z_{\rm begin}}\ ,
\end{align}
and 
\begin{align}
\frac{\nu_{\rm 21cm}}{1+z_{\rm end}-\Delta z'_2}=\frac{\nu_{\rm 21cm}+{\rm FWHM}_{\nu}}{1+z_{\rm end}}\ .
\end{align}
We adopted $\Delta z_1=10\Delta z'_1$ and $\Delta z_2=10\Delta z'_2$ in this paper to avoid artefacts.

The C21LRT calculations were done with a desktop with 8 cpu cores and the computational time of each calculation was shorter than 1 minute. In principle, for any ionisation structure which spans $z_{\rm begin}$ and $z_{\rm end}$, C21LRT can resolve down to arbitrarily small spatial scales with accuracy by adopting appropriate $\Delta z_1$ and $\Delta z_2$ and using multiple rays.

\bsp	% typesetting comment
\label{lastpage}
\end{document}